\documentclass[12pt,preprint]{aastex}

\newcommand{\vsh}{\hspace*{0.12cm}}

\slugcomment{accepted by the PASP; December 7, 2004}

\shorttitle{Carbon stars in the IRTS survey}
\shortauthors{Le~Bertre et al.}

\begin{document}


\title{Carbon stars in the IRTS survey}


\author{T.~Le~Bertre}
\affil{LERMA, UMR 8112, Observatoire de Paris, 61 av. de l'Observatoire,
           F-75014 Paris, France}
\email{Thibaut.LeBertre@obspm.fr}

\author{M.~Tanaka\altaffilmark{1}, I.~Yamamura and H.~Murakami}
\affil{Institute of Space and Astronautical Science,  3-1-1 Yoshinodai, 
           Sagamihara, Kanagawa 229-8510, Japan}
\and

\author{D.J.~MacConnell}
\affil{Science Programs, Computer Sciences Corporation, 
           Space Telescope Science Institute, 3700 San Martin Drive, 
           Baltimore, MD 21218, U.S.A.}


\altaffiltext{1}{National Astronomical Observatory, Mitaka, Tokyo, 181-8588, 
Japan}


\begin{abstract}
We have identified 139 cool carbon stars in the near-infrared 
spectro-photometric survey of the  InfraRed Telescope in Space (IRTS) 
from the conspicuous presence of molecular absorption bands at 1.8, 3.1 and 
3.8 $\mu$m. Among them 14 are new, bright (K $\sim$ 4.0--7.0), 
carbon stars. We find a trend relating the 3.1 $\mu$m band strength to the 
K$-$L' color index, which is known to correlate with mass-loss rate. 
This could be an effect of a relation between the depth of the 3.1 $\mu$m 
feature and the degree of development of the extended stellar atmosphere 
where dust starts to form.
\end{abstract}



\keywords{-- stars: carbon -- stars: mass-loss -- stars: AGB and post-AGB 
       -- infrared: stars}


\section{Introduction}

Carbon stars have traditionally been searched for in optical surveys with 
objective-prism plates. They were selected either through, initially, 
C$_2$ bands in the blue (Secchi 1868) 
or, later, through CN bands in the ``photographic 
infrared'' (Nassau \& Velghe 1964). 
More recently carbon stars have been selected from spectroscopic data 
obtained in the optical range in the course of the Sloan Digital Sky Survey 
(Margon et al. 2002).

Carbon stars have also been searched for in the IRAS survey on the basis 
of an emission feature visible at 11.3 $\mu$m in the Low Resolution Spectra
and attributed to silicon carbide (Volk \& Cohen 1989; 
Epchtein et al. 1990). A method to find carbon stars undergoing 
mass loss by combining IRAS and near-infrared (1--5 $\mu$m) color indices
has been proposed by Epchtein et al. (1987).
This method has been successfully used by Guglielmo et al. (1993) 
to discover new infrared carbon stars.

Carbon stars have been found also by selecting candidates from near-infrared 
color indices and then performing optical spectroscopy (Mauron et al. 2004).

The formation processes of our Galaxy have left traces at large distances
from the Sun which are best seen in the form of evolved stars. 
The fraction of carbon stars among evolved stars is an indicator of the 
metallicity in the original population, and there is evidence that this 
fraction is increasing with galactocentric distance (Guglielmo et al. 1993).
Inventories of cool carbon stars are of special interest because 
as these objects are evolved they 
can be used to trace matter at large galactocentric distances out to 
the Magellanic Clouds (Ibata et al. 2001). 

The latest published version of the General Catalog of Galactic Carbon Stars 
(CGCS) has 6891 entries (Alksnis et al. 2001); most of them have been found 
on objective-prism photographic plates.

In the present work, we examine the potential of finding carbon stars by using 
near-infrared spectro-photometric surveys. For that purpose we use the data 
provided by the Japanese space experiment IRTS.

\section{Near-infrared spectro-photometry with the IRTS}


The InfraRed Telescope in Space (IRTS) is a 15-cm diameter cooled telescope, 
operated in space, which surveyed, by continuous scanning, 
$\sim$\,7~\% of the sky (2700 deg.$^2$) 
with 4 infrared instruments (Murakami et al. 1996). 
The Near-InfraRed Spectrometer (NIRS) is a grating spectrometer
that covers two spectral ranges, 1.4--2.5 $\mu$m and 2.8--4.0 $\mu$m, 
with a spectral resolution ranging from $\sim$~20, at short wavelengths, 
to $\sim$~40 at the long wavelength end. The instrument was mainly designed 
to study the extended infrared emission and has an entrance aperture
of 8\,\arcmin\,$\times$\,8\,\arcmin. However, more than 14\,000 
point sources were also detected as peaks (``events'') over this 
extended emission. Their spectra are useful to probe 
the physical properties of late-type stars (e.g. Matsuura et al. 1999). 
We have also exploited them to study the contribution of mass-losing AGB stars 
to the galactic cycle of matter (Le~Bertre et al. 2001, Paper~I, and 
2003, Paper~II).

Hereafter we use the 2002 data release (NIRS Point Source Catalogue --PSC-- 
Version 1) which is described in Yamamura et al. (2003). For each event, 
a spectrum was constructed from the flux measurements in every channel 
by fitting the scan signal templates. Then the spectra corresponding 
to the same point source were identified and averaged. Errors are estimated 
from the errors on each measurement and the scatter of the data 
points{\footnote{The IRTS data and explanations are available via the DARTS 
archive; URL: {\it http://www.darts.isas.jaxa.jp/}.}}.




\section{Cool carbon stars in the NIRS Point Source Catalogue}

\begin{figure}
\begin{center}
\includegraphics[angle=-90,scale=.40]{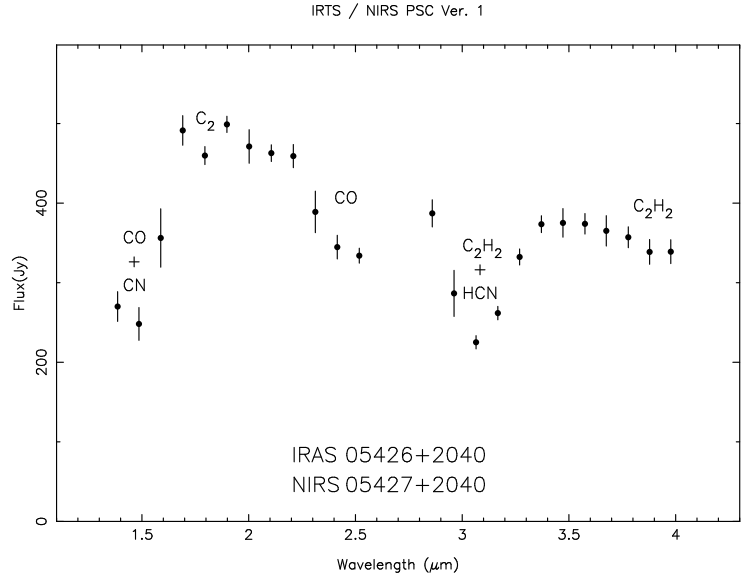}
\caption[]{NIRS spectrum of CGCS 1042 ($\equiv$ Y Tau; N2, C5II)}
\label{YTau_Ctype_wav}
\end{center}
\end{figure}

Cool carbon stars are known to show molecular absorption bands in the 
near-infrared range. A NIRS spectrum of a bright carbon star, Y Tau, is 
presented in Fig.~\ref{YTau_Ctype_wav}. A C$_2$ band at 1.8 $\mu$m,  
a band attributed to C$_2$H$_2$ at 3.8 $\mu$m and a blend at 3.1 $\mu$m, 
due to C$_2$H$_2$~+~HCN (Ridgway et al. 1978), are clearly visible  
as well as the CO bands at 1.6 and 2.3 $\mu$m. 
The C$_2$H$_2$ band at 3.8 $\mu$m may not be visible in all carbon stars: 
Yamamura et al. (1998) show that it is seen preferentially in the intermediate 
mass-losing stars of the class III defined by Groenewegen et al. (1992). 
Its presence may also depend on the phase of variability; 
for instance it is visible in all ISO Short Wavelength Spectrometer (SWS) 
spectra of V CrB except the one obtained close to maximum (see in Sect.~4.2, 
Fig.~\ref{VCrB}), possibly an effect of the strong dependence of 
the C$_2$H$_2$ abundance on the stellar effective temperature 
(Matsuura et al. 2002).
CN also contributes to the steep decline at the shortest wavelengths 
($\sim$~1.4\,$\mu$m).

\begin{figure}
\begin{center}
\includegraphics[angle=-90,scale=.40]{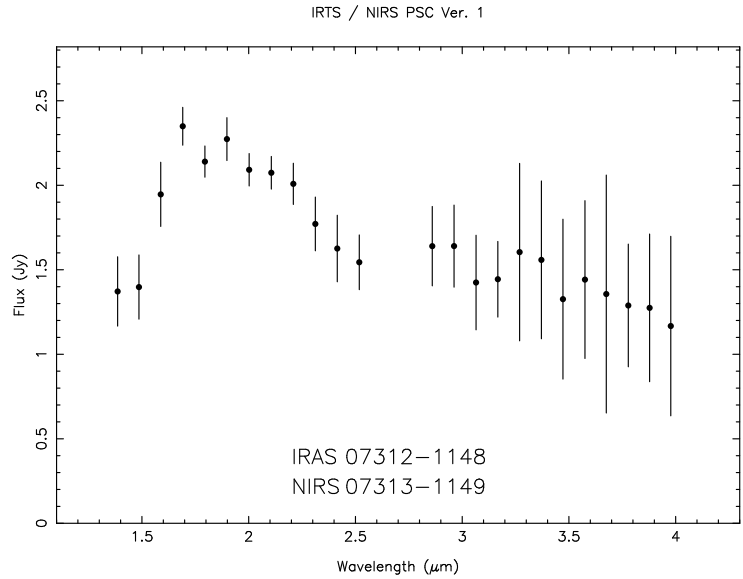}
\caption[]{NIRS spectrum of CGCS 1767 (N).}
\label{spec_S07313-1149}
\end{center}
\end{figure}

In general, the 3.1 $\mu$m blend is much deeper than the band due to C$_2$. 
However, we caution that, in the NIRS spectra, the spectral resolution around 
1.8 $\mu$m is lower than around 3.1 $\mu$m which may lead to 
underestimating the corresponding C$_2$ band depth, the C$_2$ band width 
being about 0.05 $\mu$m (Lan\c con \& Wood 2000). 
Also, as the entrance aperture 
was larger than the point sources, the individual spectra may have been 
displaced slightly over the detector array hindering the perception of this  
C$_2$ band. This band can be
useful when the signal-to-noise ratio is marginal in the range 
2.8--4.0 $\mu$m (Fig.~\ref{spec_S07313-1149}).  
Also it is reassuring to detect it as the presence of C$_2$ absorption bands 
is the defining characteristic of C stars in the 0.4--0.6 $\mu$m region.

\begin{figure}
\begin{center}
\includegraphics[angle=-90,scale=.40]{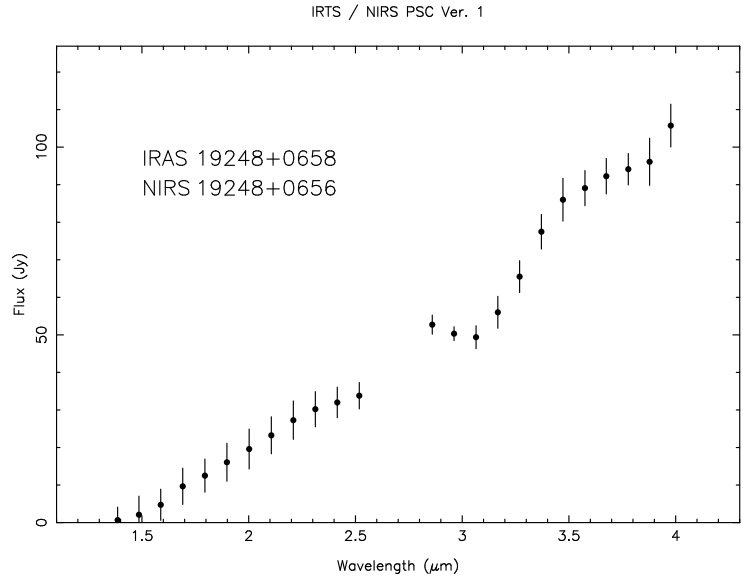}
\caption[]{NIRS spectrum of CGCS 4275 ($\equiv$ V1421 Aql, AFGL 2392).}
\label{spec_S19248+0656}
\end{center}
\end{figure}

The 3.1 $\mu$m blend is seen even in highly reddened sources 
(Fig.~\ref{spec_S19248+0656}) although it tends to be obliterated by 
circumstellar dust emission.

To identify carbon stars in the NIRS PSC, we use the 3.1 $\mu$m index defined 
in Paper II. This index gives the depth of the band (at 3.1 $\mu$m) 
relative to the continuum (taken at 2.9 and 3.3 $\mu$m); 
the maximum depth is obtained for an index value of 0.0 whereas a value 
$\sim$ 0.5 indicates a flat spectrum.
The strength of the 3.1 $\mu$m feature has been found to be a very 
sensitive indicator of the C/O abundance ratio (Catchpole \& Whitelock 1985).

\begin{figure}
\begin{center}
\includegraphics[angle=-90,scale=.40]{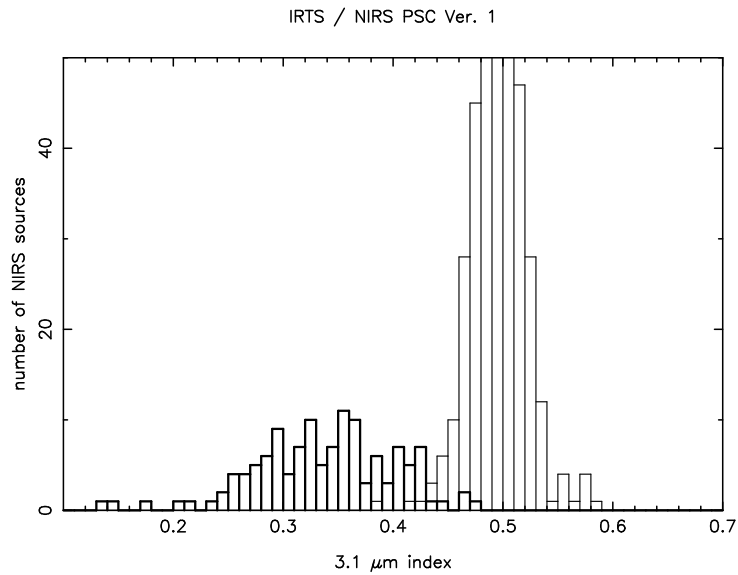}
\caption[]{Histogram of the 3.1 $\mu$m index for the sample of AGB 
mass-losing stars studied in Paper II. Thick line: 126 C-rich stars, 
thin line : 563 O-rich stars. For clarity the histogram has been truncated.}
\label{histog_XFC_KL_CO}
\end{center}
\end{figure}

In Paper II we have used this index to discriminate oxygen-rich stars from 
carbon-rich ones in a sample of 689 mass-losing AGB stars. A histogram of 
this sample is presented in Fig.~\ref{histog_XFC_KL_CO}. Most of the carbon 
stars have an index smaller than 0.45 (only 3 sources have an index larger 
than 0.45); the median value is at 0.34. On the other hand, the oxygen-rich 
stars have in general an index around 0.5. 

\begin{figure}
\begin{center}
\includegraphics[angle=-90,scale=.40]{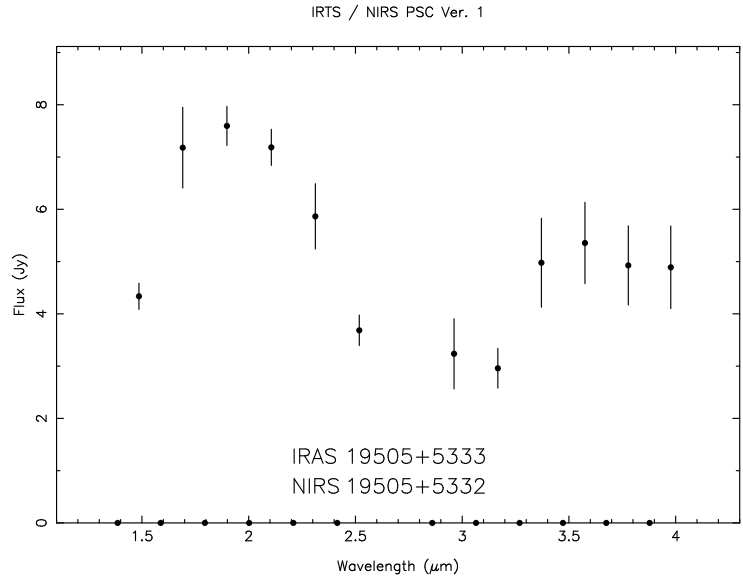}
\caption[]{NIRS spectrum of BS Cyg ($\equiv$ CGCS 4543; Nep, SC8/8).}
\label{BSCyg}
\end{center}
\end{figure} 

We should note at this stage that there is, in principle, 
a difficulty with some 
S stars with a C/O abundance ratio close to 1, and in particular the SC stars. 
If such S/SC stars show the 3.1 $\mu$m absorption band, with our procedure, 
they would be recognized as C stars, although strictly speaking they are 
only somewhat enriched in carbon.  The SC class refers to stars that 
show ZrO bands 
and also CN bands. Some show the 3.1 $\mu$m absorption band 
(Catchpole \& Whitelock 1985) as well as a band 
attributed to CS around 4 $\mu$m (Aoki et al. 1998). 
They constitute a rare group (only 14 SC stars are in the General Catalog of 
S Stars -- Stephenson 1984) forming a continuous sequence from S to C stars.
In principle, they do not show C$_2$ bands (Stephenson 1984). As most of our 
sources show the C$_2$ band at 1.8 $\mu$m and as SC stars are rare, 
we do not expect a significant contamination. On the other hand, the 
discovery of more such rare objects would be of interest. 
In the IRTS survey, we found only one source 
(NIRS\,19505+5332\,$\equiv$\,BS\,Cyg\,$\equiv$\,CGCS\,4543) which belongs to 
this class. Its spectrum, for which only the odd-numbered channels were 
transmitted, shows clearly the absorption band at 3.1 $\mu$m 
(Fig.~\ref{BSCyg}). It is classified as SC8/8 by Keenan \& Boeshaar (1980)  
but does not appear in Stephenson (1984). 
Catchpole \& Whitelock (1985) note that the SC star BH Cru which has 
an absorption feature at 3.1 $\mu$m may show occasionally characteristics of a 
CS star (Lloyd Evans 1985). 
This is supported by the recent discussion of this star 
by Zijlstra et al. (2004) who interpret such changes as a temperature effect.

\begin{figure}
\begin{center}
\includegraphics[angle=-90,scale=.40]{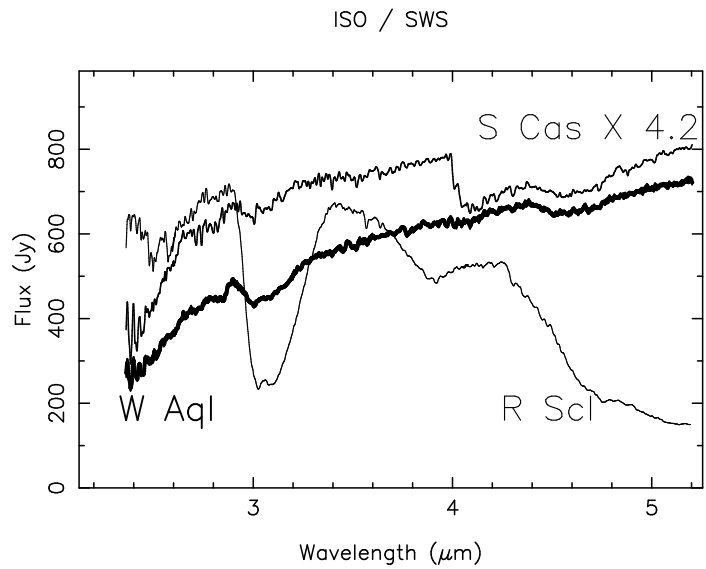}
\caption[]{ISO/ SWS spectra of two S-type stars, W Aql (CSS~1115, thick line) 
and S Cas (CSS 28, medium line), and one C-type star, R Scl (CGCS 234, thin 
line). A factor of 4.2 has been applied to the spectrum of S Cas.}
\label{ISOSWS_WAql_SCas_RScl}
\end{center}
\end{figure}

Finally we note that some S stars show a weak absorption band at 3.1 $\mu$m.
Two such cases have been observed by ISO with the SWS: 
W Aql and S Cas. The near-infrared part of their spectra  
is shown in Fig.~\ref{ISOSWS_WAql_SCas_RScl} together with the one of 
a carbon star, R Scl. The 3.1 $\mu$m blend is visible clearly
in R Scl and W Aql and seems also present in S Cas. The latter source has
been observed by the IRTS, but the 3.1 $\mu$m band is not visible
in the NIRS spectrum (see Fig.~4 in Paper I). The source W Aql
is not in the NIRS PSC; the 3.1 $\mu$m index measured on 
the ISO spectrum is $\sim$ 0.46. On the other hand,  
the C$_2$H$_2$ band at 3.8 $\mu$m is present in the spectrum of R Scl but not 
in the spectra of the two S stars. There is an absorption from 
4.0 to 4.2 $\mu$m due to SiO in the spectrum of S Cas and perhaps 
W Aql but not in the one of R Scl. 
At the present stage, we conclude that some S-type stars may show the 
3.1 $\mu$m absorption band, and we suspect that they do not show the band of 
C$_2$H$_2$ at 3.8 $\mu$m. For these stars, the 3.1 $\mu$m seems to stay weak, 
and we expect that a limit of 0.45 on the index value keeps them out of 
our selection.

We have considered the 8879 NIRS sources for which the data are of 
sufficiently good quality to estimate their K and L' magnitudes 
in the ESO photometric system (see Paper II). 
Among them we obtained 471 sources with a 3.1 $\mu$m index $\leq$ 0.45. Each 
corresponding spectrum was examined visually. We found 139 sources that  
unambiguously show a C$_2$ band at 1.8 $\mu$m and/or the blend at 3.1 $\mu$m  
and 220 which are very probably not carbon-rich. The former are reported in 
Table~\ref{tab_C-rich}.  
Most of these carbon stars 
can be associated with an IRAS and a CGCS source. However, these associations 
should be viewed with caution because the entrance aperture of the NIRS on 
the sky was 8\arcmin$\times$8\arcmin. Also, the accuracy of the pointing 
reconstruction is $\sim$ 2\arcmin ~r.m.s. 

Some carbon stars may have been missed, 
for instance among the remaining 112 sources for which the NIRS spectra 
do not have a signal-to-noise ratio sufficient to characterize them. Also,  
we remind that some carbon stars may have a 3.1 $\mu$m index above 0.45\,:
three such cases were identified in Paper II (see also 
Fig.~\ref{histog_XFC_KL_CO}). For clarity, we have added 
these 3 cases at the end of Table~\ref{tab_C-rich}. 
The majority of the 220 sources that we have rejected have 
a 3.1 $\mu$m index in the range 0.40--0.45. Also, there are 
sources which show an emission band at 3.3 $\mu$m attributed to PAHs. 
This emission raises the flux at 3.3 $\mu$m relative to 3.1 $\mu$m which 
mimics an absorption. These ``PAH'' sources, which often have a large K$-$L', 
are readily identified from their NIRS spectra. Finally, it is worth noting 
that we did not find carbon stars with a K$-$L' color smaller than 0.4.

Tanaka et al. (in preparation) have developed a similar, but slightly 
different, method. They use 5 parameters to separate M, C and S stars. 
Applying this method to 4002 NIRS spectra of good quality, they find 91 carbon 
stars. With our procedure we have obtained all their carbon stars, but one 
(NIRS 07281$-$1216 $\equiv$ IRAS 07280$-$1217 $\equiv$ CGCS 1735). This 
source shows a weak 3.1 $\mu$m absorption band (3.1 $\mu$m index = 0.46) 
which explains why we could not select it. For completeness 
we have added this source at the end of Table~\ref{tab_C-rich}.
The incidence of carbon stars in their sample is about 0.023 (91/4002),  
whereas it is about 0.016 (139/8879) in our sample. This confirms 
that we miss some carbon stars at low signal-to-noise ratio. 

The few sources (14) in Table~\ref{tab_C-rich}  
that have no CGCS counterpart can be considered as 
new cool carbon stars. Their NIRS spectra are given in Fig.~\ref{NIRS_new_CS}.
It is surprising that new carbon stars are found that do not have an extreme 
color (K$-$L' $\leq$ 1) 
and still are relatively bright (K $\sim$ 6). 
Perhaps they have been missed in previous surveys
because they are variable. Some of these new carbon stars are in fields 
covered by near-IR objective-prism plates obtained by one of us 
(MacConnell 2003). These plates have been re-examined, and the results 
are given in the notes to Table~\ref{tab_C-rich}. 

\section{Discussion}

\subsection{Cool carbon stars}\label{CoolCS_discuss}

\begin{figure}
\begin{center}
\includegraphics[angle=-90,scale=.40]{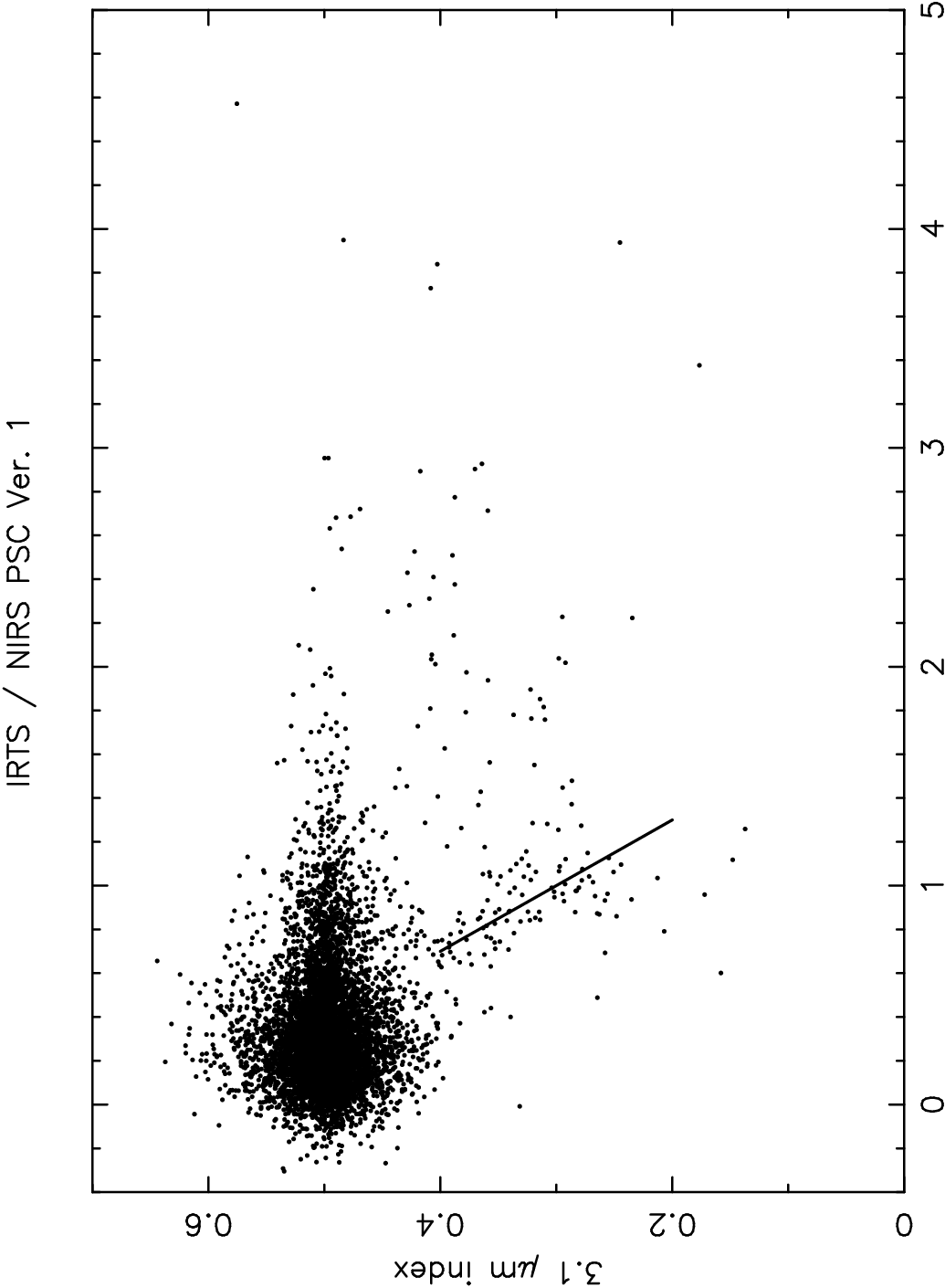}
\includegraphics[angle=-90,scale=.40]{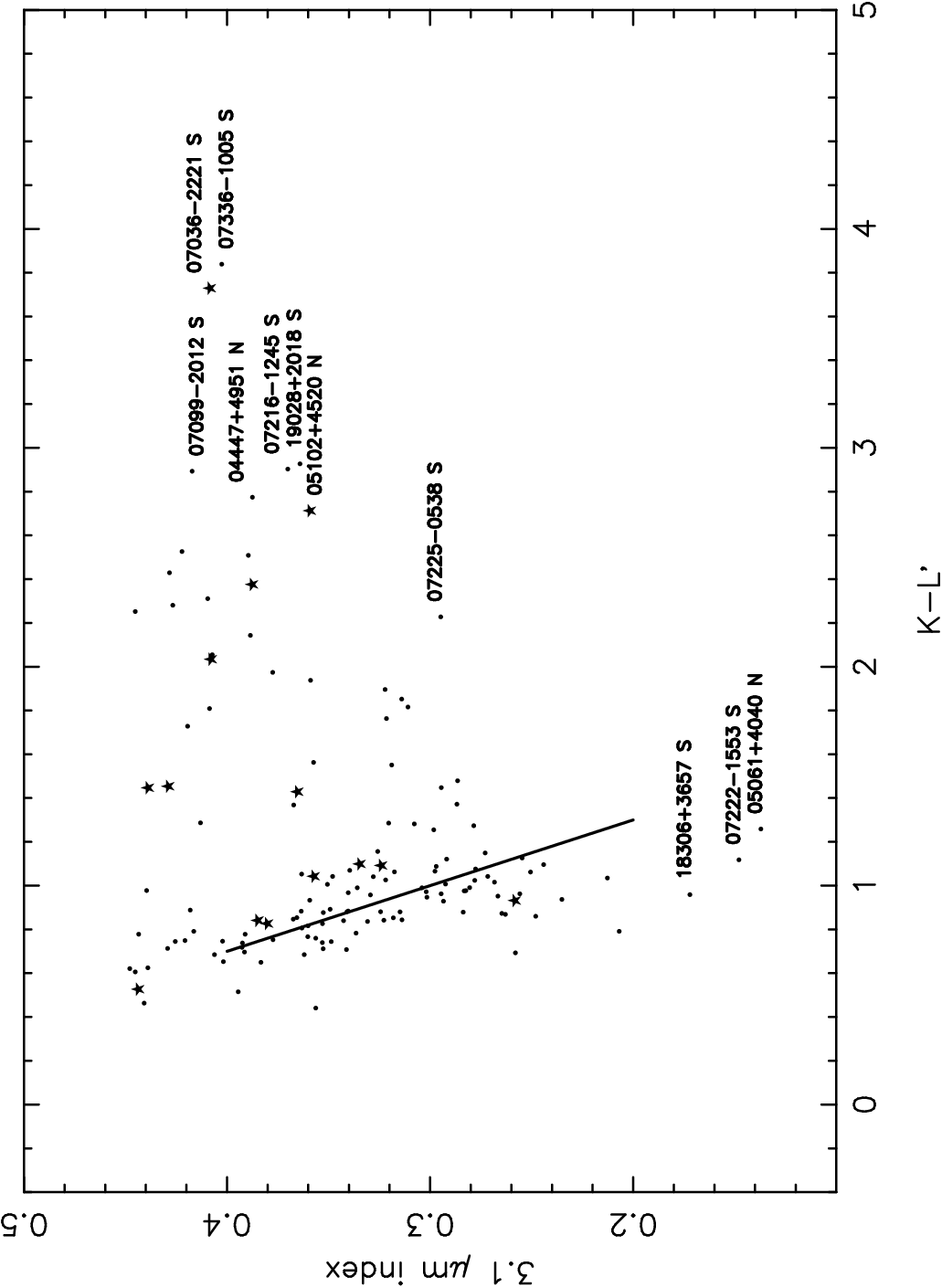}
\caption[]{3.1 $\mu$m index versus K$-$L'. Upper panel: All NIRS sources 
(8879). Lower panel: carbon-rich sources (139); the most extreme sources 
are marked by their NIRS identifiers and the 14 new carbon stars are 
represented by a $\star$.}
\label{diag_C_KminL}
\end{center}
\end{figure}

In the upper panel of 
Fig.~\ref{diag_C_KminL} we present the 3.1 $\mu$m index as a function of 
K$-$L'. Most of the 8879 NIRS sources with available K and L' magnitudes 
are in a large clump around an index value 
of $\sim$~0.50 and a K$-$L' color of $\sim$~0.3. There is a tail going 
horizontally to the right and composed mainly of oxygen-rich late-type stars. 
In addition, there is a ``plume'' going downward to the right,  
which is marked by a straight line in the figure. This plume is made of  
carbon stars as can be seen in the lower panel of the figure.
 
This feature shows that there is a relation between the 3.1 $\mu$m index that 
characterizes the carbon-rich stellar atmospheres and the K$-$L' color 
in the range 0.7 to 1.4. This color is 
known to be related to the present mass-loss rate of carbon stars (Le Bertre 
1997). It is interesting to note that the same kind of plume is visible when 
plotting the 1.9  $\mu$m index (which characterizes O-rich late-type stars) 
versus K$-$L' which, for these sources, characterizes 
their present mass-loss rate (Le Bertre \& Winters 1998). 
The trend is not a perfect correlation and sources are found also to the 
right of the plume. This might be an effect of dust emission filling in
the 3.1 $\mu$m band in spectra of sources undergoing heavy mass loss. 
Groenewegen et al. (1994) have obtained 2.8--3.5 $\mu$m spectra for 
16 carbon stars with a K$-$L index in the range 2--7. They find that the 
3.1 $\mu$m feature weakens with redder K$-$L color and interpret 
this tendency also as an effect of filling-in by dust emission.

A possible interpretation of the trend observed in Fig.~\ref{diag_C_KminL} is 
that the mass loss and the absorption band at 3.1 $\mu$m are both increasing 
with the degree of development of the extended stellar atmosphere where 
dust is likely to form. Winters et al. (2000) have shown that the mass loss 
of red giants is favored by a low effective temperature, a low stellar mass, 
a high luminosity and a large pulsation amplitude. The synthetic spectra of 
carbon-rich AGB star atmospheres obtained by Loidl et al. (1999) from dynamic 
models show that the strengths of the C$_2$H$_2$ and HCN features increase 
in the same manner, because these molecules form preferentially 
at low temperature (T\,$<$\,2500\,K) in the upper atmospheric layers.

We have found 14 new carbon stars in a sample of 139 sources. 
The C$_2$ band at 1.8 $\mu$m has been useful to confirm the carbon-rich nature 
of the sources in the cases for which the signal-to-noise ratio at 3 $\mu$m 
is poor. Unfortunately, the spectral resolution of the NIRS around 2 $\mu$m 
($\sim$~20) is not enough to define an index measuring the depth of this band. 

The IRTS survey is limited in sensitivity to K $\sim$ 8 and L' $\sim$ 7; 
furthermore in some regions of the sky, close to the Galactic Plane where 
the level of the extended emission is increased by confusion, 
the sensitivity was reduced by $\sim$ 2 magnitudes (Paper~II). 
Most of the carbon stars that we have identified are probably at 
a distance between 1 and 5 kpc from the Sun (see Figure 7 in Paper II).
Extrapolating to the whole sky, a survey like the IRTS would reveal 
$\sim$ 2000 cool carbon stars with possibly 200 new ones. These 
estimates would rise greatly for a survey with better sensitivity, spatial 
resolution and spectral resolution.

The advantages of searching for cool carbon stars by near-infrared 
spectro-photometric means are multiple. The effective temperature of 
N-type stars ranges from 1800 to 4000 K (Bergeat et al. 2001), and 
they reach the maximum of their energy distribution around 2 $\mu$m. Many of 
them are undergoing mass loss and are surrounded by circumstellar dust shells 
which shift their spectra towards long wavelengths ($\sim$ 3 $\mu$m or more). 
As a result of the energy distribution, the infrared data provides also a more 
direct determination of the luminosity than those obtained in the optical. 
Furthermore, carbon stars are often variable with an amplitude which decreases 
with wavelength; therefore, it is less likely to miss a star that is at  
minimum if the survey is done at a long wavelength. 
There is also a general advantage of the infrared range for detecting sources 
in regions of the Galaxy which are affected by dust extinction  
(``zone of avoidance''). 
Finally, as cool carbon-star spectra are characterized by wide molecular 
bands, a spectral resolution of $\ge$ 40 is enough to clearly recognize them.
With spectro-photometry it is easy to build photometric indices targeted 
to specific classes of sources. On the other hand, 
our experience (Sect. 3) shows that 
the availability of a continuous spectrum over a large wavelength range 
is useful to identify the sources that contaminate the 
samples which are extracted on the basis of these photometric indices: 
the same data are used to select candidates and to confirm them. 

\subsection{Other carbon stars in the IRTS survey}\label{OtherCS_discuss}

There are other categories of carbon stars, the majority 
being found in the R and CH classes. The R stars are a mix of giants
with effective temperature lower than $\sim$ 5000 K, in general non-variable,
and of classical cool carbon stars (late R type/ N type). The R stars proper 
(excluding N type) are too faint to be on the AGB and do not show 
evidence of mass loss. They may owe their carbon enrichment to an He-core 
flash (Dominy 1984) and, perhaps, to a coalescence in a	binary system (McClure 1997). 
They are more easily selected by isolating the early R0--R4 sub-types.

\begin{figure}
\begin{center}
\includegraphics[angle=-90,scale=.40]{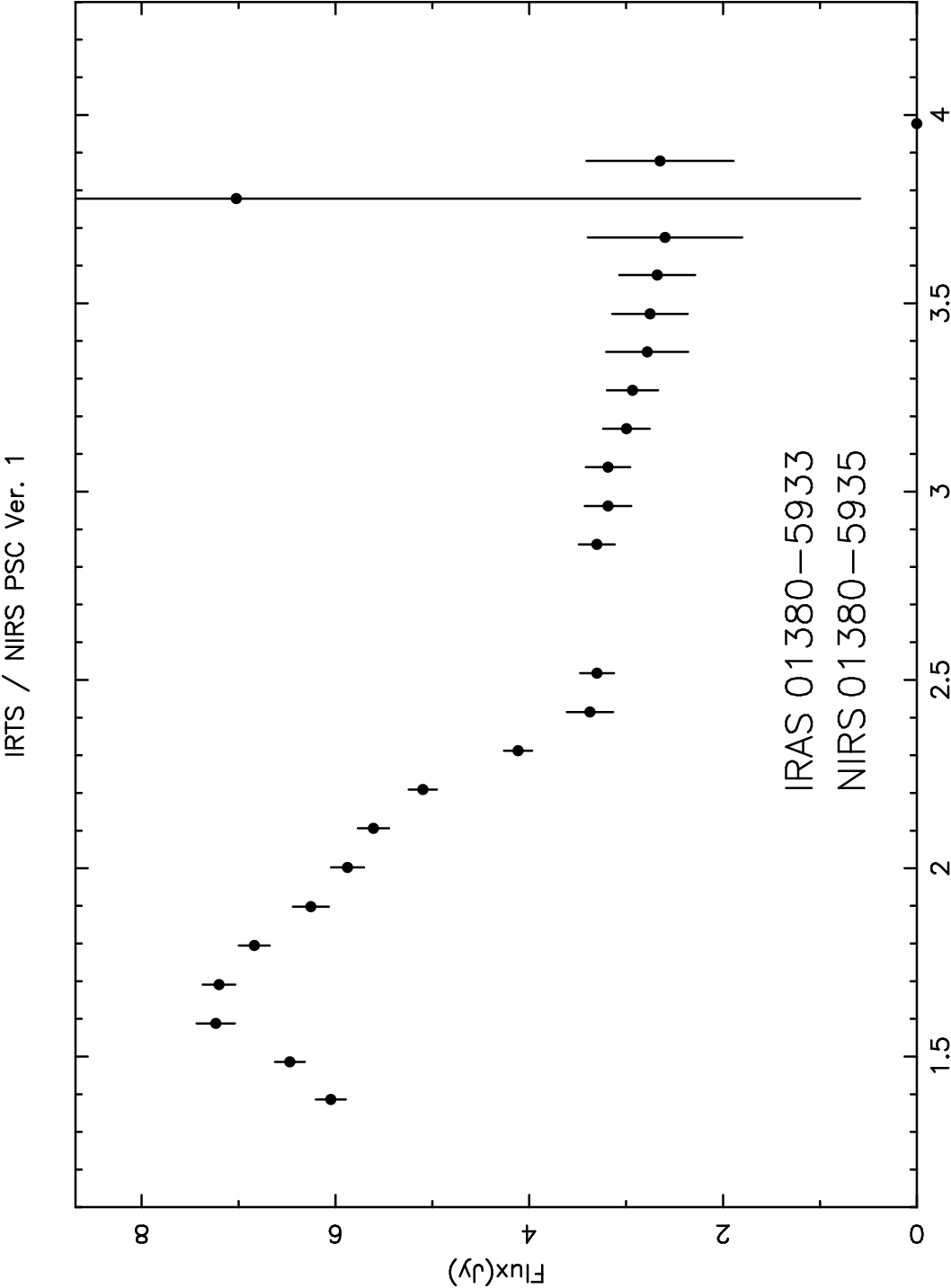}
\includegraphics[angle=-90,scale=.40]{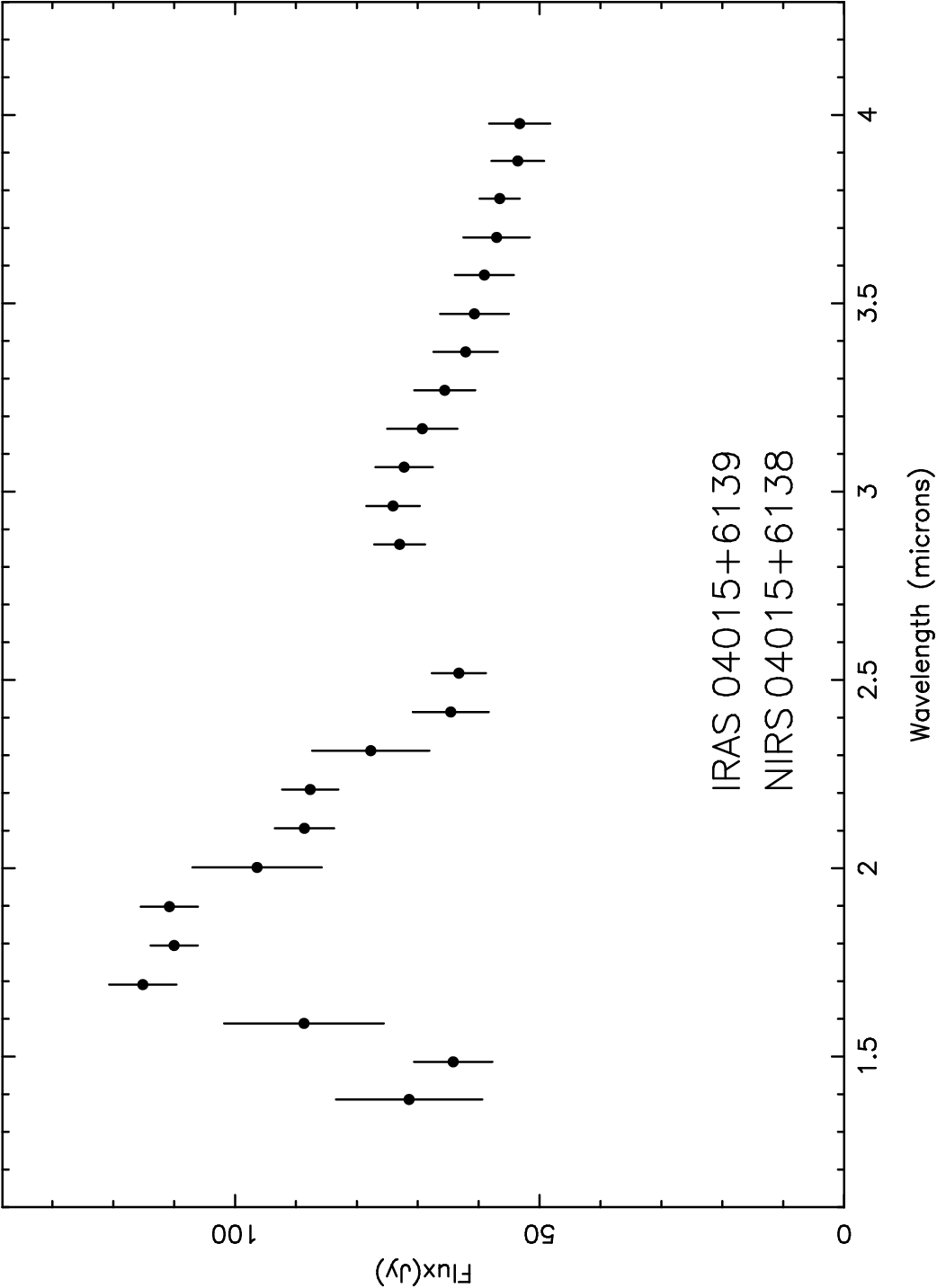}
\caption[]{NIRS spectra of two R-type stars. Upper panel: CGCS~77 (R3). 
Lower panel:  CGCS~177 (UV Cam, R8).}
\label{Rtype}
\end{center}
\end{figure}

A search for R stars in the NIRS PSC was performed by cross-correlating the 
positions in the PSC with those in the CGCS catalogue (Alksnis et al. 2001).
We found several associations. The NIRS spectra of sources associated with 
early (R0--R4) type stars show sometimes CO absorptions at 1.6 and 2.3 $\mu$m, 
but no other distinctive features (Fig.~\ref{Rtype}, upper panel). R stars of 
later type may show the C$_2$ band at 1.8 $\mu$m without the 3.1 $\mu$m 
absorption band (Fig.~\ref{Rtype}, lower panel), confirming 
our suspicion that the latter is not sufficient to select all cool carbon stars. 
It thus appears again that the 1.8 $\mu$m band, as well as the 3.1 $\mu$m band, 
could be very useful for future surveys. 

\begin{figure}
\begin{center}
\includegraphics[angle=-90,scale=.40]{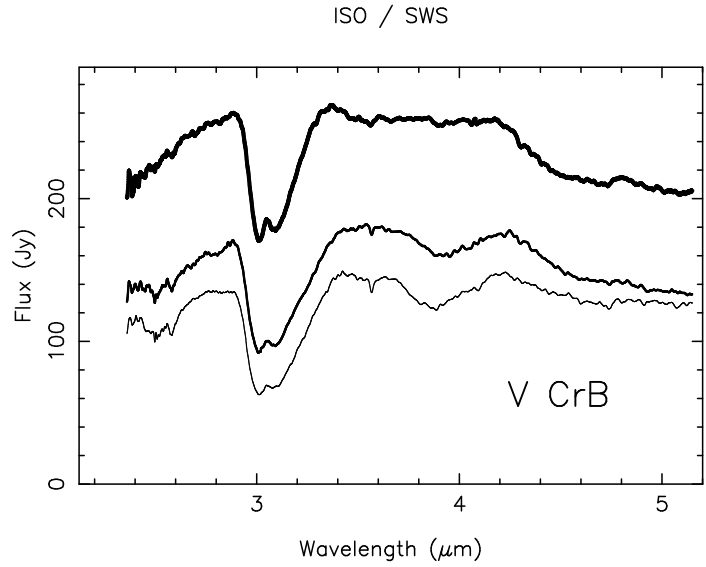}
\caption[]{ISO SWS spectra of a CH star (V CrB, CGCS~3652) obtained 
at 3 different phases. Note that the 3.8 $\mu$m absorption is virtually 
absent (as well as the 3.56 $\mu$m HCN band) close to maximum (thick line), 
but clearly present close to minimum (medium and thin lines).}
\label{VCrB}
\end{center}
\end{figure}

The galactic CH stars are metal-poor carbon stars which are characterized 
in the optical range by strong CH bands. They are spectroscopic binaries 
and probably owe their carbon enrichment to mass transfer from a more evolved 
companion (Jorissen 1999). We have not identified any CH star in the NIRS PSC. 
However, at least 3 members of this class (V~Ari, V~CrB, HD~189711) 
were observed by ISO with the SWS. Apart from the long wavelength wing of 
the 2.3 $\mu$m CO band, no characteristic feature is visible in the 
(2.3--4.1 $\mu$m) spectra of V~Ari and HD~189711; unfortunately, the region of 
the spectrum around 1.8 $\mu$m is missing in the ISO spectra. V~CrB shows 
a strong absorption at 3.1 $\mu$m and at some phases an absorption 
at 3.8 $\mu$m (Fig.~\ref{VCrB}). However, this source is a Mira variable 
and therefore probably not typical of the CH class. If objects like V~CrB
are in the NIRS PSC, they would be selected as cool carbon stars and 
probably not recognized as being of the CH-type. 

There are also the J-type stars which are characterized by strong $^{13}$C 
bands indicating a low ($\leq$ 10) $^{12}$C/$^{13}$C ratio. We have 
identified three such cases which were already selected in Sect. 3: 
NIRS~07235$-$0354 (CGCS~1708), NIRS~12545+6615 (CGCS~3313, RY~Dra) 
and NIRS~18306+3657 (CGCS~4038, T~Lyr). All show strong molecular bands 
at 1.8 and 3.1 $\mu$m. At the low spectral resolution of the NIRS, it does  
not seem possible to isolate the J-type stars from the other cool carbon stars. 

\begin{figure}
\begin{center}
\includegraphics[angle=-90,scale=.40]{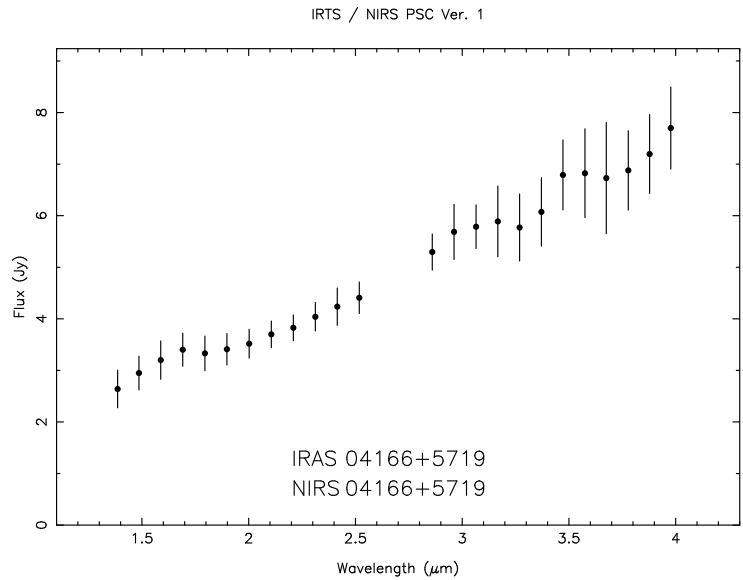}
\caption[]{NIRS spectrum of TW Cam, a carbon-rich RV Tauri star (RVa; F8Ib).}
\label{TWCam}
\end{center}
\end{figure}

We found 3 NIRS sources which could be associated with RV Tauri stars
(Table~\ref{tableother}). It is known that some RV Tauri show a carbon-rich 
nature. 
Giridhar et al. (2000) have performed an abundance analysis of several 
RV Tauri stars: U Mon is O-rich (C/O $\sim$ 0.8) whereas TW Cam is carbon-rich
(C/O $\sim$ 2). The O-rich character of U Mon is supported by its 
IRAS Low Resolution Spectrum (LRS class 26). Le Bertre et al. (2001) note 
that, in a K$-$L' versus 12$-$25\,$\mu$m diagram, 
TW Cam falls in the 'c' (carbon-rich) 
region of Epchtein et al. (1987). The NIRS spectra corresponding to TW~Cam 
(Fig.~\ref{TWCam}) and U~Mon are basically featureless, whereas 
the one of NIRS\,18125+0510 shows an H$_2$O band at 1.9 $\mu$m and a CO band 
at 2.3 $\mu$m. The latter NIRS source is certainly not carbon-rich.

\begin{table}
\begin{center}
\caption{Other candidate stars.\label{tableother}}
\begin{tabular}{lccllll}
\tableline\tableline
NIRS name    & K & K$-$L' & IRAS association & other ass. & Spec. type &
 reference \\
\tableline
04166$+$5719 N & 5.57 & 1.70 & 04166$+$5719  & TW Cam     & F8Ib       & 
Simbad \\
07284$-$0940 S & 3.61 & 1.33 & 07284$-$0940  & U Mon      & K0Ibpvar   & 
Simbad \\
18125$+$0510 N & 5.19 & 1.07 & 18123$+$0511  &            & G5         & 
Mass (2003) \\
\tableline
\end{tabular}
\end{center}
\end{table}

Finally we could not identify RCB stars in the NIRS PSC, 
nor any known carbon-rich post-AGB stars. 
At the present stage, we do not see an easy way to pinpoint warm carbon stars 
directly in the IRTS survey.

\section{Conclusion}

We have identified 139 cool carbon stars in the IRTS/NIRS survey 
on the basis of the 3.1 $\mu$m absorption feature due to a blend 
of C$_2$H$_2$ and HCN bands. About 90\% have been confirmed through 
cross-identification with the CGCS and 14 are new, relatively 
bright (K $\sim$ 3.9--6.9), C stars. This indicates that an important 
number of bright carbon stars are still missing from our inventories. 

It is likely that there are more cool carbon stars in the Point Source 
Catalogue that were not selected either because the 3.1 $\mu$m band is weak 
or because a limited signal-to-noise ratio hampered an unambiguous 
classification. The spectral resolution of the IRTS was insufficient 
to fully exploit the important C$_2$ band at 1.8 $\mu$m; 
a spectral resolution of 40 or better would have been preferable. 
The C$_2$ band at 1.8 $\mu$m and the C$_2$H$_2$ band at 3.8 $\mu$m should be 
useful to separate C from S stars. An extension of the wavelength coverage 
beyond 4 $\mu$m to cover the SiO band would also have been helpful.

Sensitive near-infrared spectro-photometric surveys covering the range from
$\sim$~1.5 to $\sim$~4.5 $\mu$m with a spectral resolution
$\geq$ 40  have therefore the potential to reveal new carbon stars in 
the Galaxy and in its satellites, as well as in other galaxies, 
and to provide useful information on their physical properties. 



\acknowledgments

We thank Dr. N. Epchtein for useful advice and encouragement. 
This research has made use of the SIMBAD database, operated at CDS, 
Strasbourg, France. We thank an anonymous referee for helpful 
suggestions.



Facilities: \facility{IRTS(NIRS)}, \facility{ISO(SWS)}.

\clearpage





\appendix

\section{Appendix material}

\begin{figure}
\begin{center}
\includegraphics[angle=-90,scale=.30]{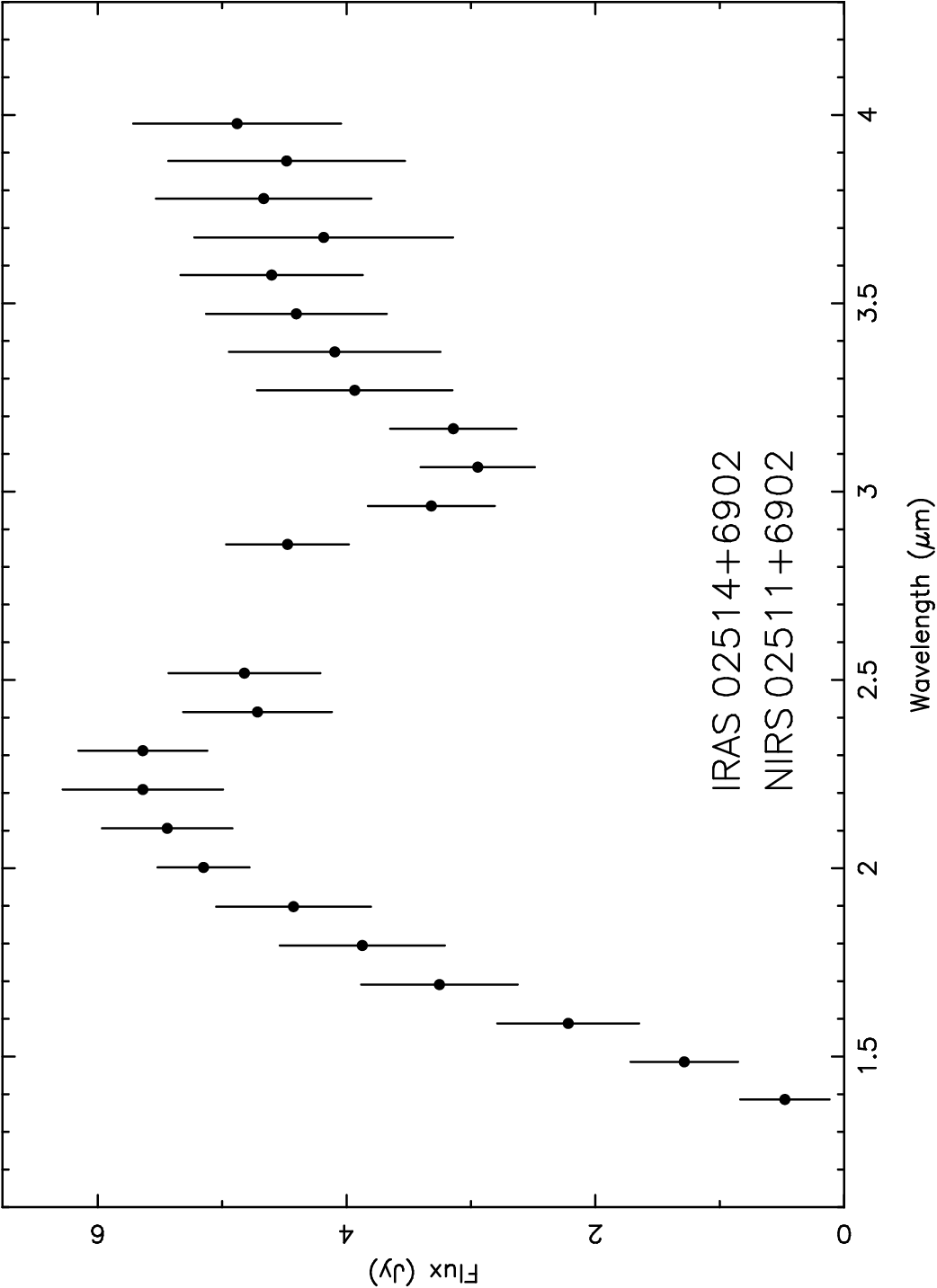}
\includegraphics[angle=-90,scale=.30]{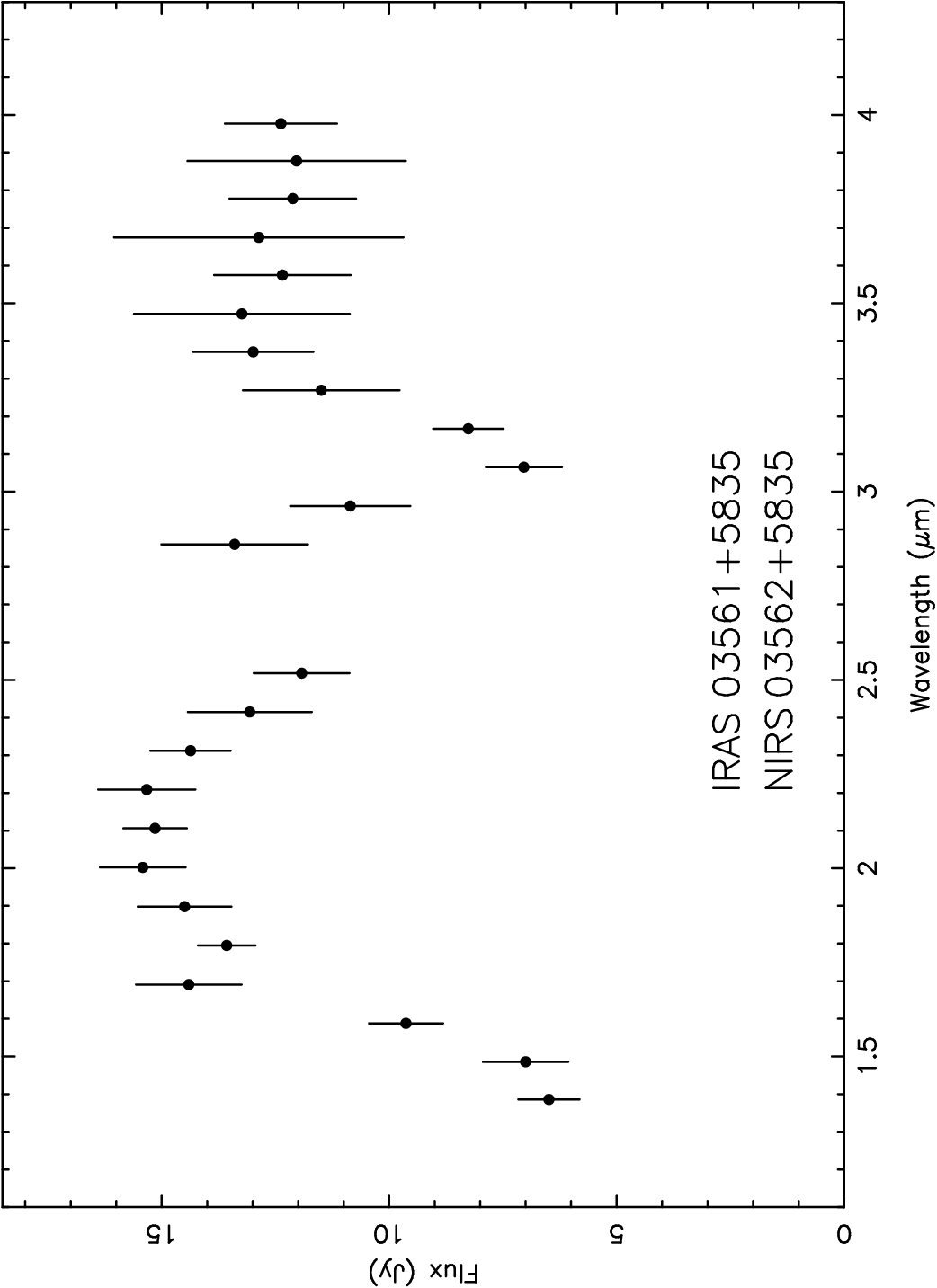}
\includegraphics[angle=-90,scale=.30]{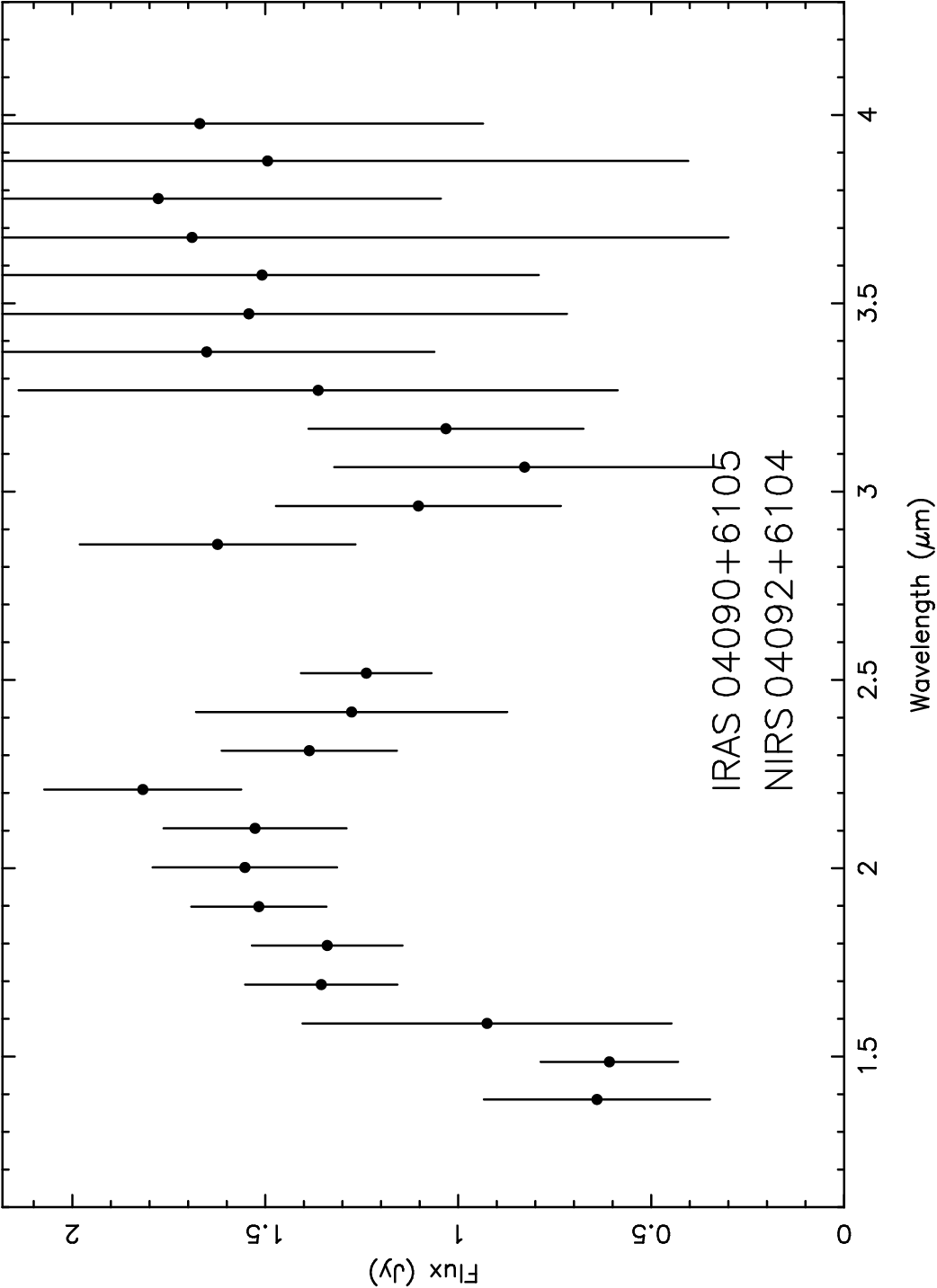}
\includegraphics[angle=-90,scale=.30]{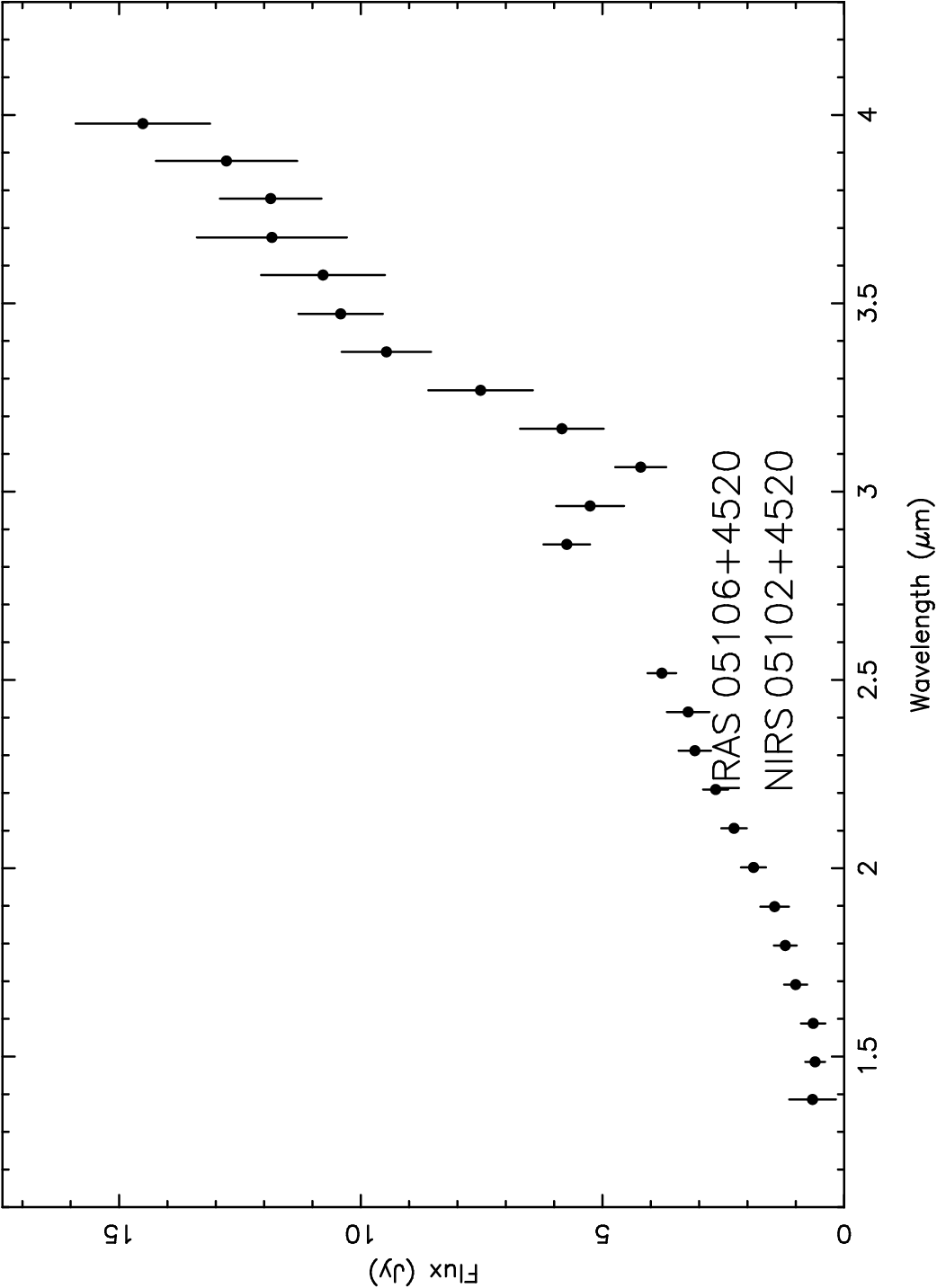}
\includegraphics[angle=-90,scale=.30]{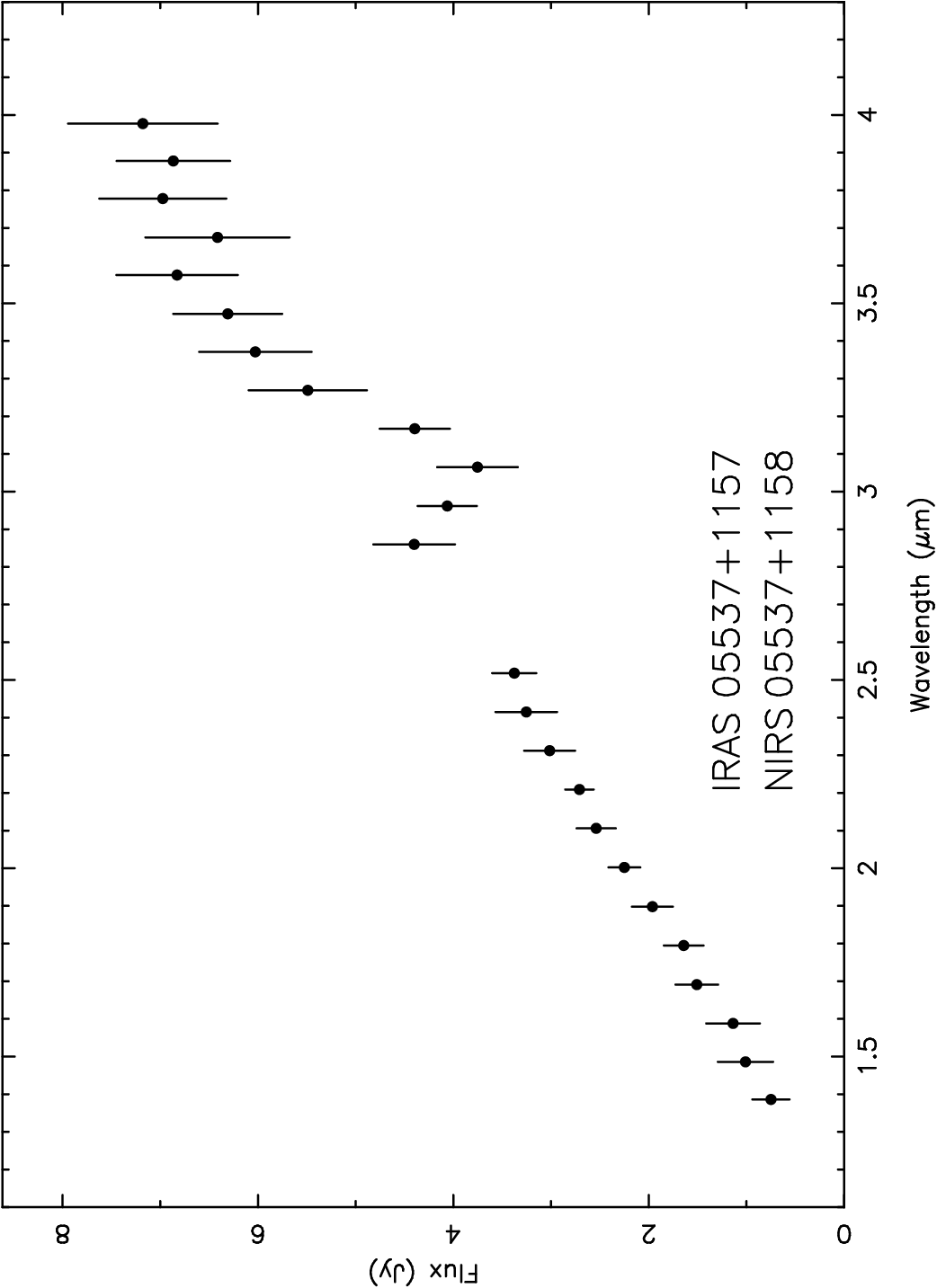}
\includegraphics[angle=-90,scale=.30]{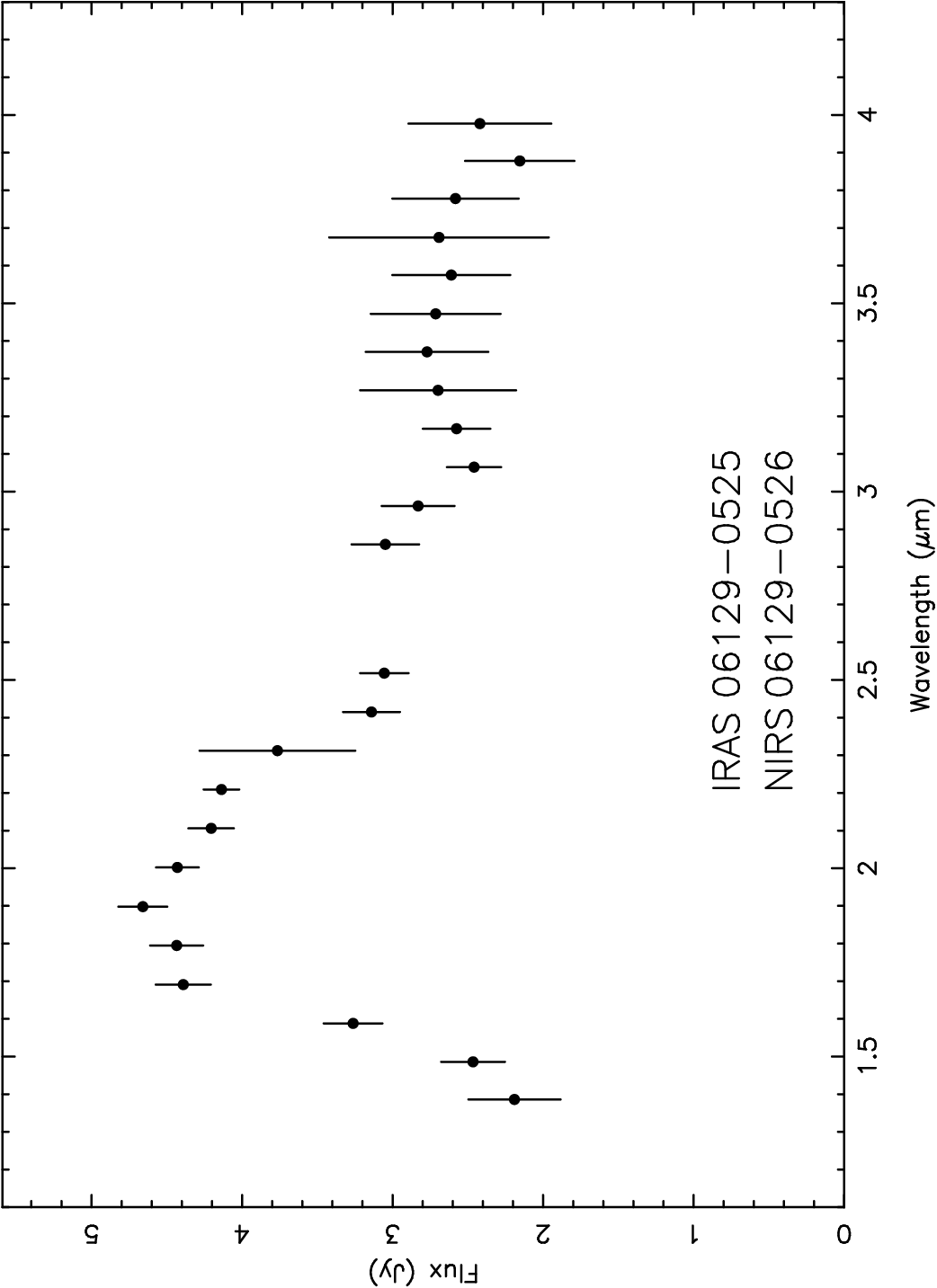}
\includegraphics[angle=-90,scale=.30]{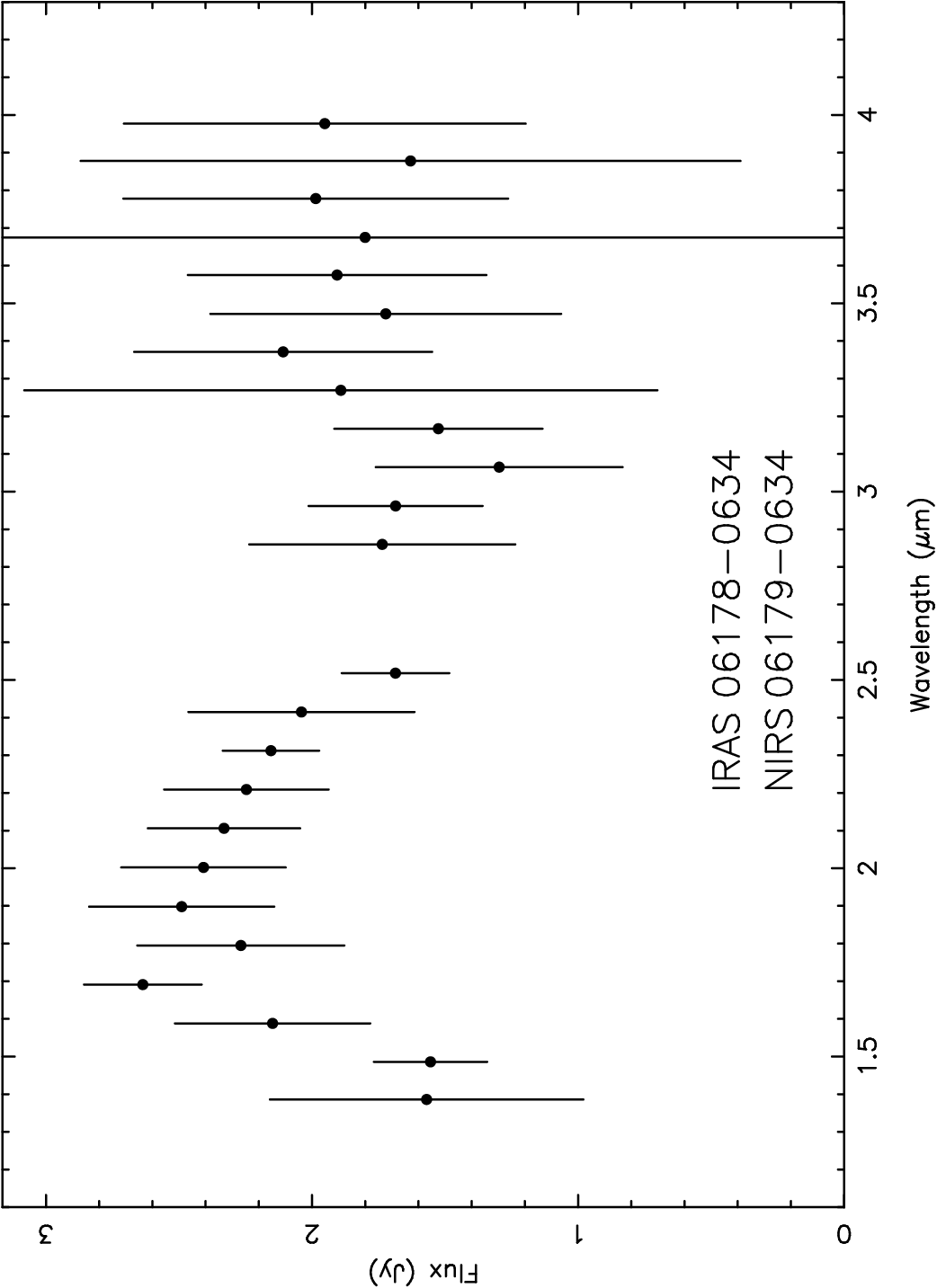}
\includegraphics[angle=-90,scale=.30]{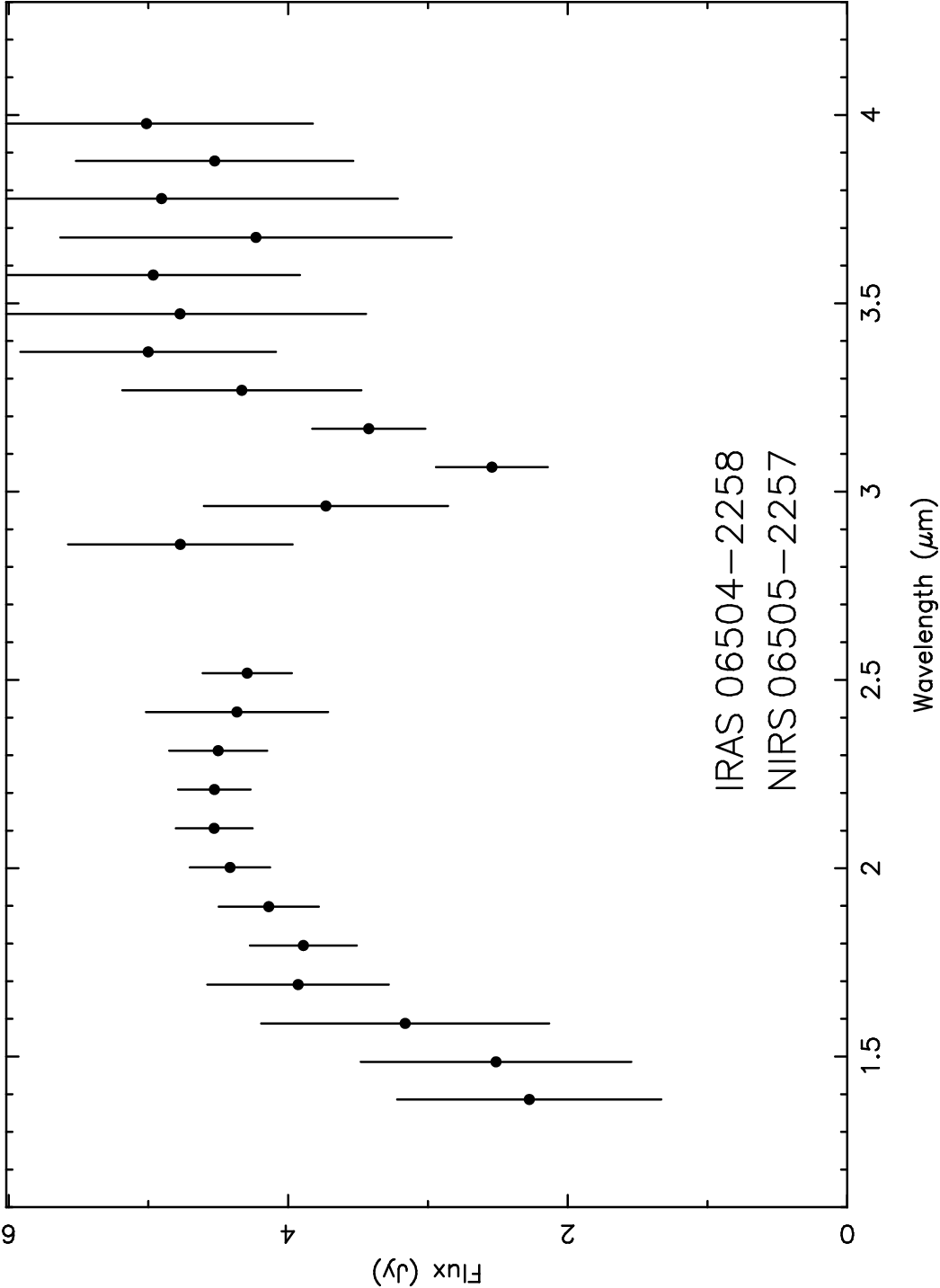}
\caption[]{NIRS spectra of new carbon stars.}
\label{NIRS_new_CS}
\end{center}
\end{figure}

\setcounter{figure}{10}
\begin{figure}
\begin{center}
\includegraphics[angle=-90,scale=.30]{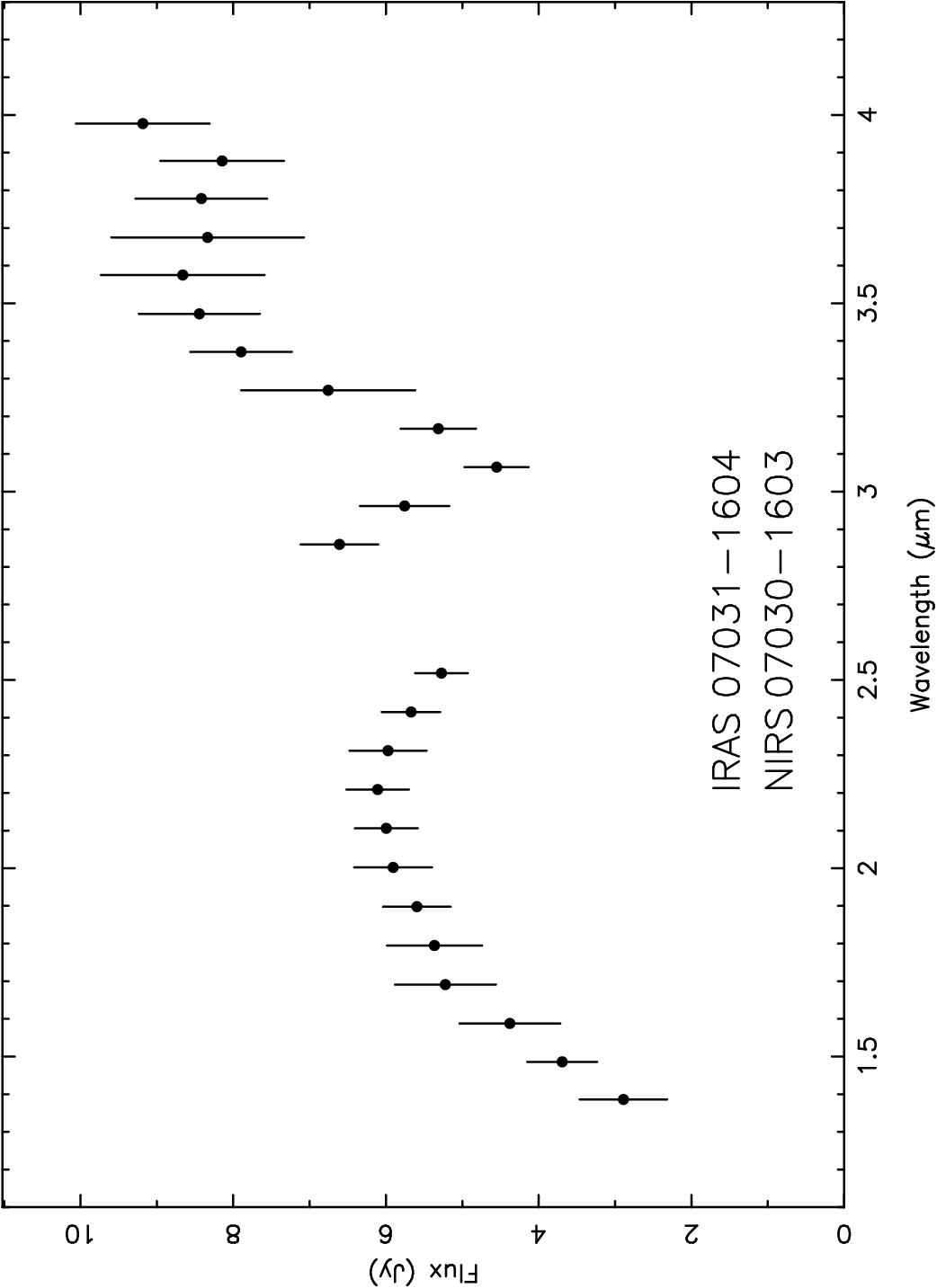}
\includegraphics[angle=-90,scale=.30]{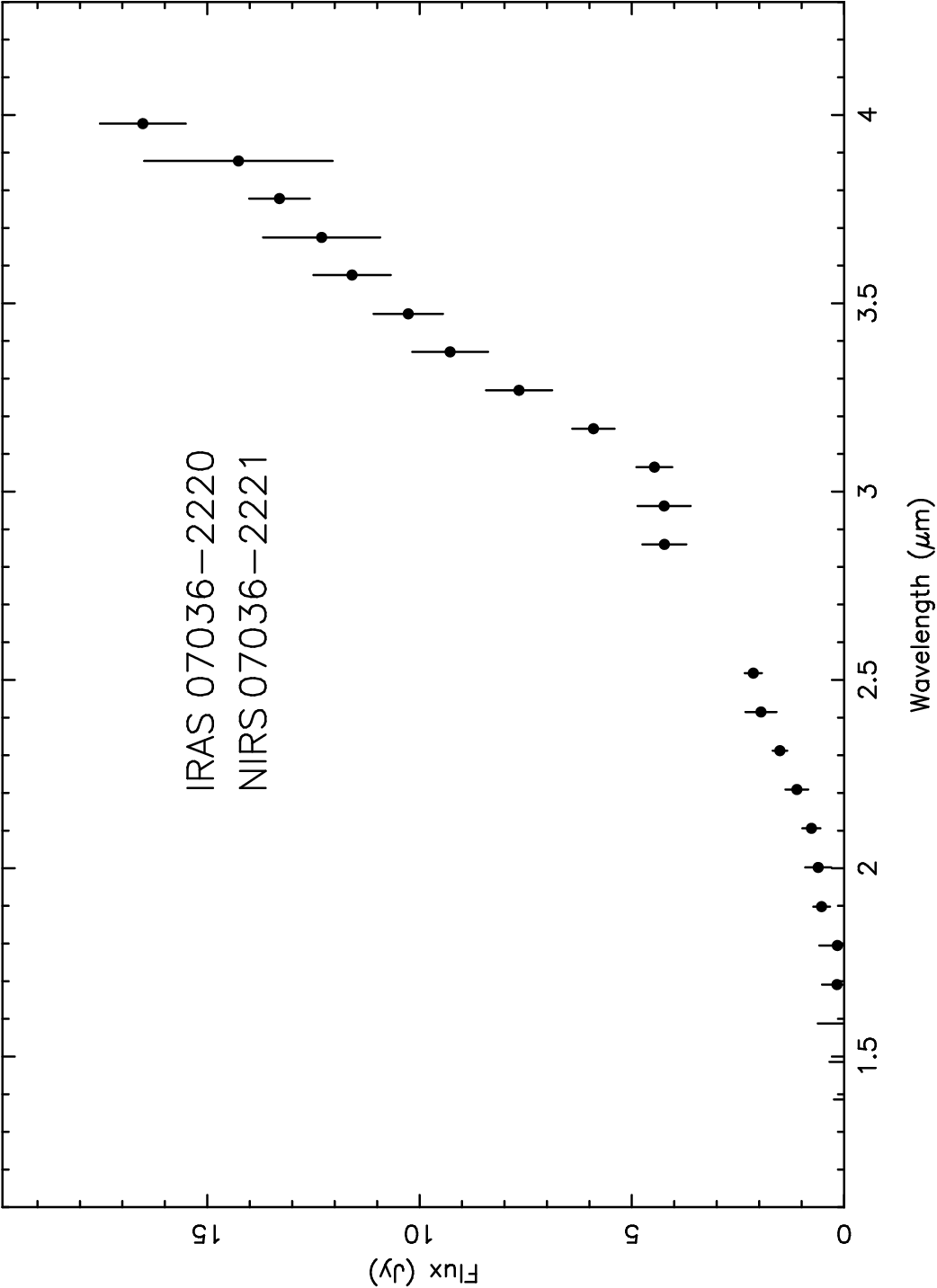}
\includegraphics[angle=-90,scale=.30]{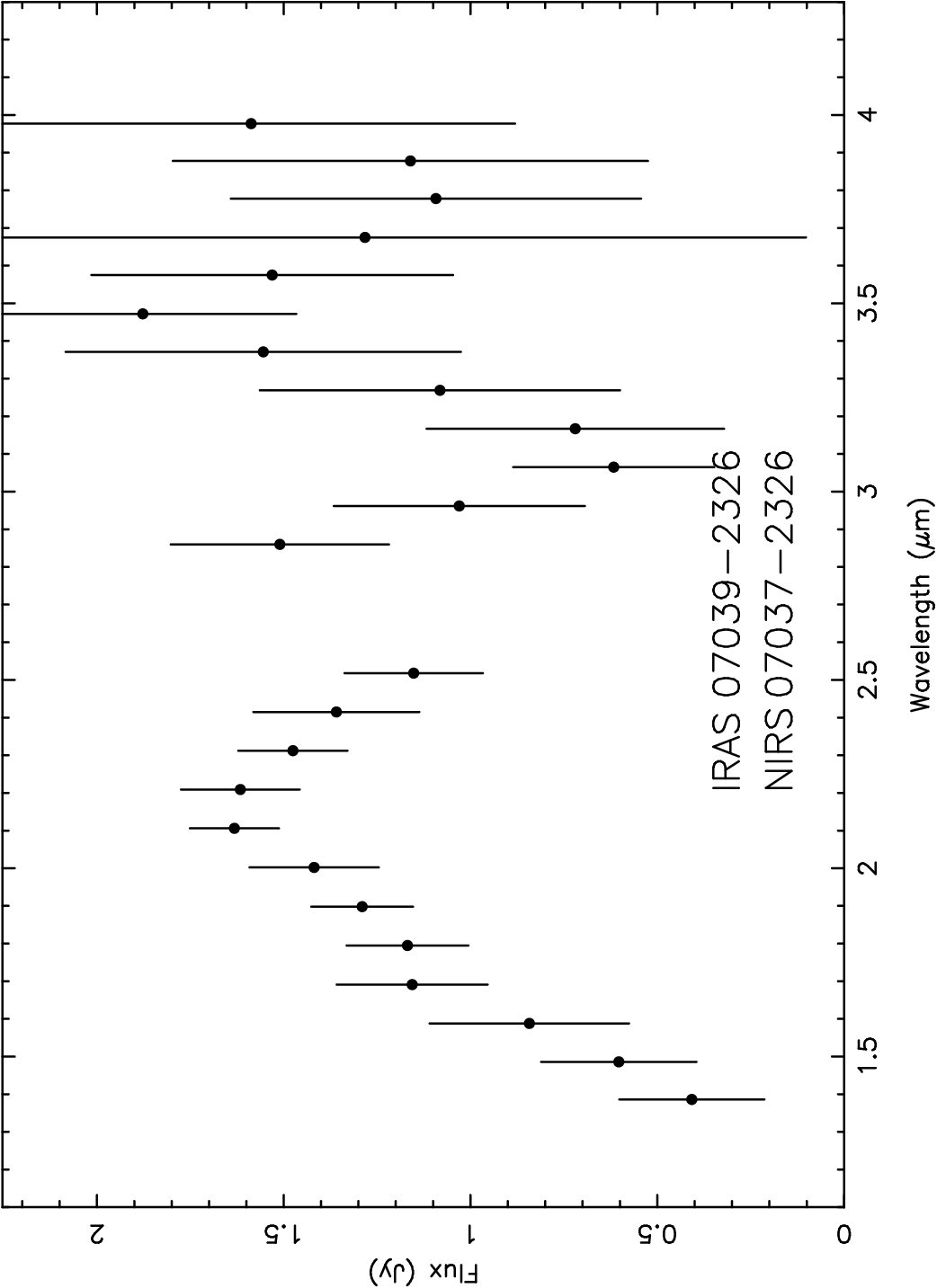}
\includegraphics[angle=-90,scale=.30]{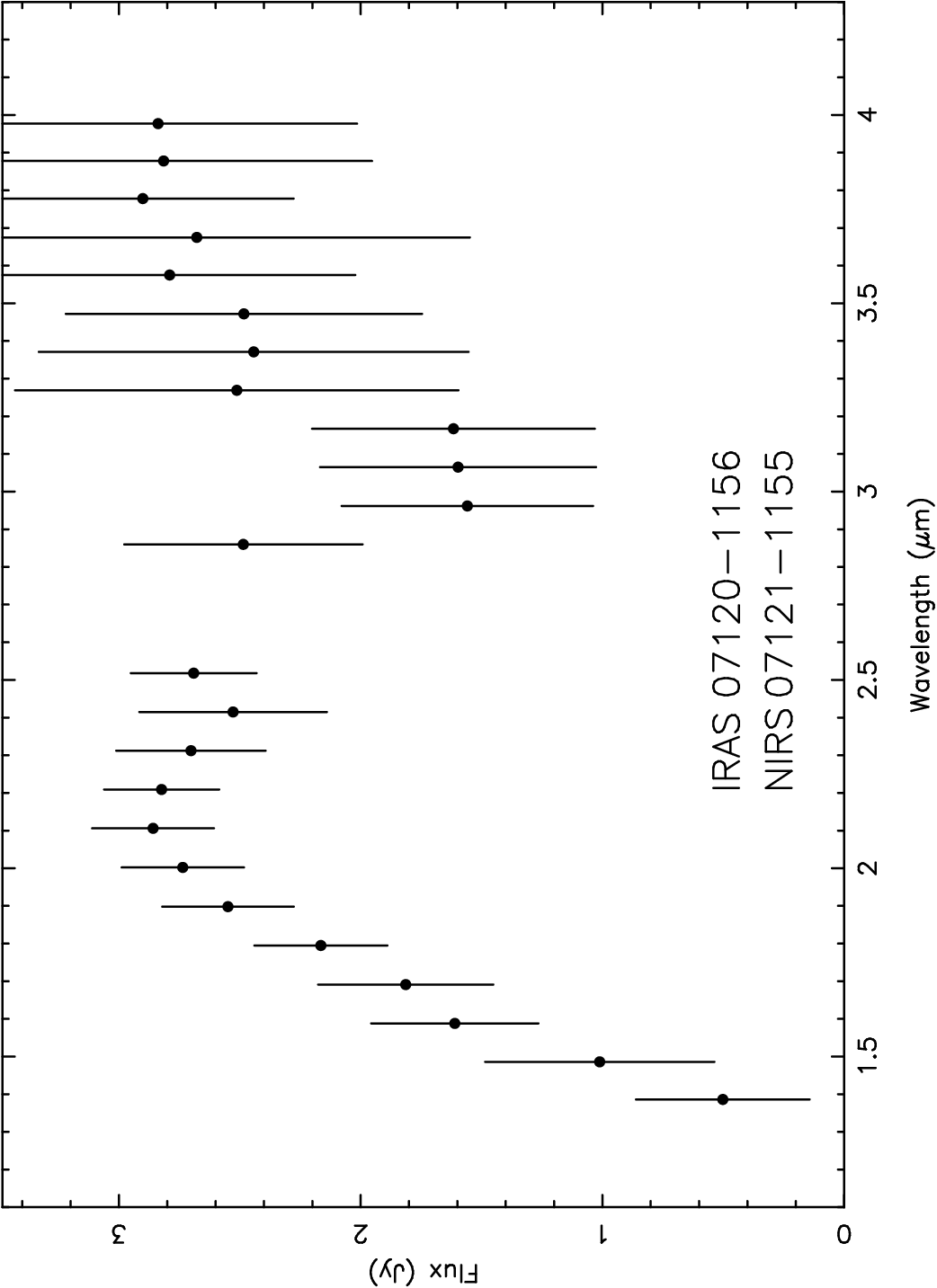}
\includegraphics[angle=-90,scale=.30]{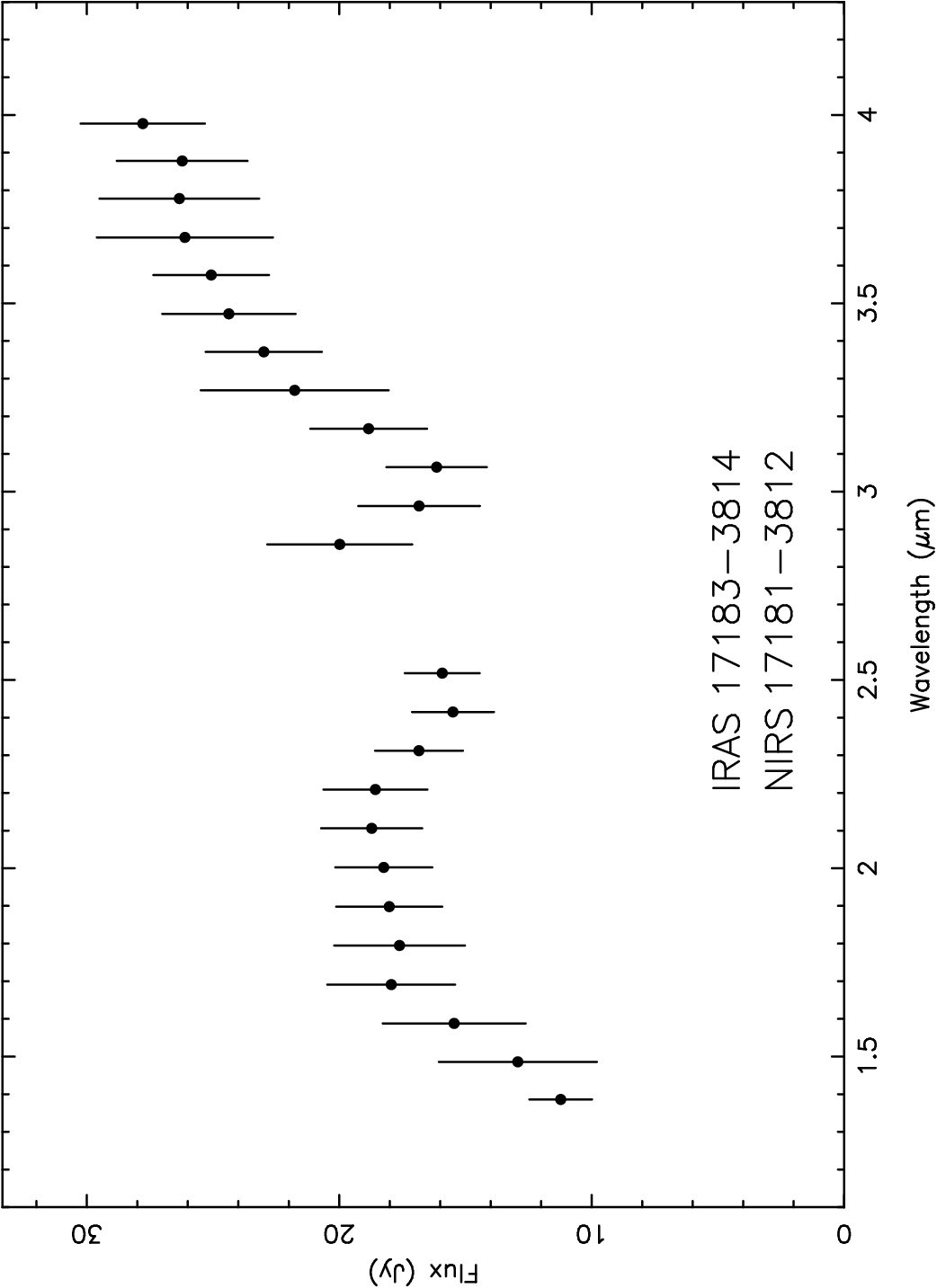}
\includegraphics[angle=-90,scale=.30]{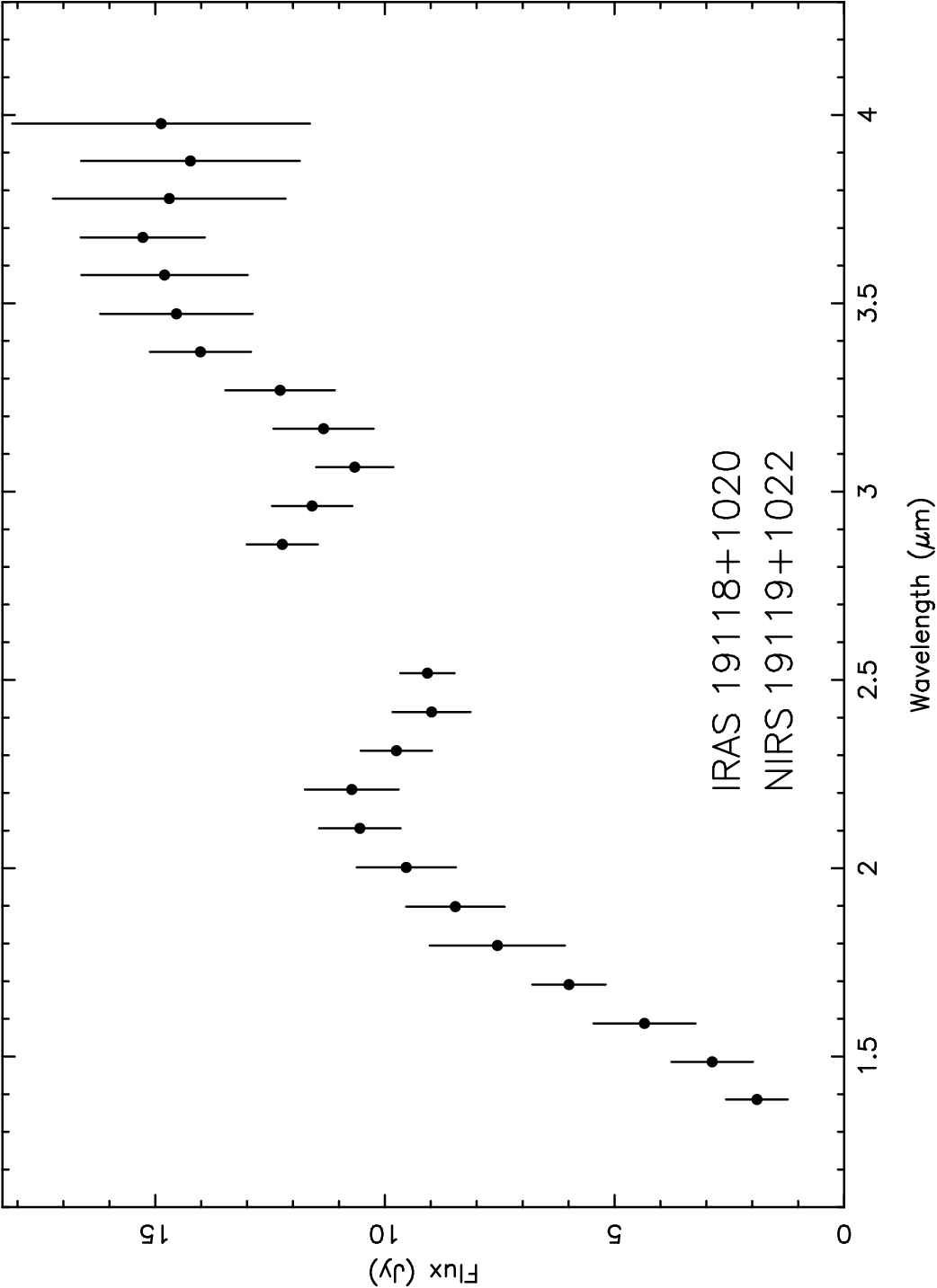}
\caption[]{NIRS spectra of new carbon stars (continued).}
\label{NIRS_new_CS}
\end{center}
\end{figure}









\clearpage

\begin{deluxetable}{lrrccclll}
\tabletypesize{\scriptsize}
\tablecaption{IRTS carbon-stars.\label{tab_C-rich}}
\tablewidth{0pt}
\tablehead{
\colhead{NIRS name} & \colhead{l$^{\rm II}$} & \colhead{b$^{\rm II}$} & 
\colhead{K} & \colhead{K$-$L'} &
\colhead{3.1 $\mu$m} & \colhead{IRAS} & \colhead{CGCS} &
\colhead{comments}
}
\startdata
 00593+7135 N  &  123.83  &    9.01  &  4.77  &  0.69  &  0.41  &  00591+7133  &  CGCS  \vsh163   &    \\
 01414+7104 N  &  127.27  &    8.91  &  4.14  &  1.37  &  0.29  &  01411+7104  &  CGCS  \vsh261   &    \\
 02511+6902 N  &  133.60  &    8.98  &  5.18  &  0.84  &  0.38  &  02514+6902  &              &    \\
 02596+6638 N  &  135.45  &    7.25  &  4.56  &  2.28  &  0.43  &  02596+6639  &  CGCS  \vsh430   &    \\
 03063+6552 N  &  136.43  &    6.91  &  5.83  &  1.03  &  0.21  &  03063+6553  &  CGCS 6044   &    \\
 03171+6515 N  &  137.71  &    6.98  &  4.71  &  0.75  &  0.43  &  03170+6515  &  CGCS 6050   &    \\
 03285+6318 N  &  139.83  &    6.03  &  6.78  &  1.01  &  0.35  &              &  CGCS  \vsh509   &    \\
 03375+6229 N  &  141.15  &    5.98  &  0.48  &  0.84  &  0.33  &  03374+6229  &  CGCS  \vsh540   &    \\
 03562+5835 N  &  145.40  &    4.37  &  4.10  &  0.84  &  0.31  &  03561+5835  &  CGCS  \vsh583   &    \\
 03562+5911 N  &  145.01  &    4.82  &  6.11  &  2.38  &  0.39  &  03561+5912  &              &    \\
 03576+5923 N  &  145.02  &    5.10  &  4.23  &  1.45  &  0.29  &  03575+5922  &  CGCS  \vsh591   &    \\
 04035+5929 N  &  145.51  &    5.66  &  5.15  &  1.29  &  0.41  &  04035+5929  &  CGCS  \vsh613   &    \\
 04092+6104 N  &  144.95  &    7.30  &  6.56  &  1.09  &  0.32  &  04090+6105  &              &    \\
 04165+5703 N  &  148.44  &    5.06  &  5.36  &  0.88  &  0.34  &  04164+5701  &  CGCS  \vsh656   &    \\
 04207+5549 N  &  149.72  &    4.59  &  4.13  &  1.04  &  0.33  &  04207+5548  &  CGCS  \vsh674   &    \\
 04217+5401 N  &  151.11  &    3.43  &  5.59  &  1.08  &  0.28  &  04216+5401  &  CGCS  \vsh677   &    \\
 04296+5253 N  &  152.75  &    3.50  &  5.28  &  0.79  &  0.42  &  04295+5252  &  CGCS  \vsh713   &    \\
 04354+5414 N  &  152.33  &    5.04  &  5.00  &  0.96  &  0.26  &  04354+5415  &  CGCS  \vsh730   &    \\
 04393+4905 N  &  156.59  &    2.07  &  5.76  &  2.14  &  0.39  &  04391+4904  &  CGCS 6085   &    \\
 04426+4950 N  &  156.37  &    2.97  &  4.68  &  1.56  &  0.36  &  04427+4951  &  CGCS  \vsh763   &    \\
 04447+4951 N  &  156.58  &    3.24  &  4.66  &  2.77  &  0.39  &  04449+4951  &  CGCS  \vsh772   &    \\
 04468+4930 N  &  157.07  &    3.27  &  5.14  &  0.94  &  0.24  &  04468+4930  &  CGCS  \vsh784   &    \\
 04500+4802 N  &  158.54  &    2.74  &  5.38  &  1.28  &  0.31  &  04500+4802  &  CGCS  \vsh801   &    \\
 04504+4950 N  &  157.19  &    3.94  &  2.85  &  1.12  &  0.29  &  04504+4949  &  CGCS  \vsh806   &    \\
 04540+4450 N  &  161.45  &    1.24  &  6.11  &  0.87  &  0.26  &  04539+4450  &  CGCS  \vsh817   &    \\
 04553+4925 N  &  158.02  &    4.28  &  5.35  &  0.88  &  0.34  &  04552+4925  &  CGCS  \vsh820   &    \\
 05058+3855 N  &  167.47  &   $-$0.65  &  2.39  &  0.71  &  0.43  &  05056+3856  &  CGCS  \vsh860   &    \\
 05061+3956 N  &  166.69  &    0.00  &  5.11  &  1.82  &  0.31  &  05061+3956  &  CGCS  \vsh862   &    \\
 05061+4040 N  &  166.11  &    0.45  &  5.72  &  1.26  &  0.14  &  05062+4040  &  CGCS  \vsh863   &    \\
 05068+3945 N  &  166.92  &    0.00  &  4.27  &  0.96  &  0.29  &  05068+3945  &  CGCS  \vsh867   &    \\
 05102+4520 N  &  162.81  &    3.82  &  5.98  &  2.71  &  0.36  &  05106+4520  &                  &    \\
 05151+3511 N  &  171.57  &   $-$1.36  &  3.61  &  1.97  &  0.38  &  05149+3511  &  CGCS  \vsh900   &    \\
 05158+3545 N  &  171.20  &   $-$0.92  &  2.72  &  0.65  &  0.38  &  05158+3544  &  CGCS  \vsh904   &    \\
 05209+3430 N  &  172.81  &   $-$0.78  &  5.58  &  1.13  &  0.25  &  05209+3430  &  CGCS  \vsh921   &    \\
 05213+4001 N  &  168.32  &    2.42  &  4.25  &  1.48  &  0.29  &  05214+4001  &  CGCS  \vsh924   &    \\
 05237+3405 N  &  173.49  &   $-$0.54  &  1.50  &  1.55  &  0.32  &  05238+3406  &  CGCS  \vsh941   &    \\
 05276+3437 N  &  173.50  &    0.43  &  4.80  &  1.04  &  0.35  &  05278+3437  &  CGCS  \vsh961   &    \\
 05278+3151 N  &  175.82  &   $-$1.06  &  4.41  &  0.88  &  0.35  &  05278+3152  &  CGCS  \vsh962   &    \\
 05285+2814 N  &  178.92  &   $-$2.94  &  5.06  &  0.75  &  0.38  &  05285+2814  &  CGCS  \vsh964   &    \\
 05287+2930 N  &  177.90  &   $-$2.20  &  4.67  &  0.85  &  0.37  &  05288+2929  &  CGCS  \vsh966   &    \\
 05292+2938 N  &  177.84  &   $-$2.03  &  3.89  &  0.97  &  0.34  &  05292+2938  &  CGCS  \vsh971   &    \\
 05334+3057 N  &  177.23  &   $-$0.56  &  5.24  &  1.81  &  0.41  &  05334+3057  &  CGCS  \vsh999   &    \\
 05346+3001 N  &  178.16  &   $-$0.84  &  3.36  &  1.03  &  0.32  &  05345+3002  &  CGCS 1003   &    \\
 05352+2247 N  &  184.34  &   $-$4.64  &  3.10  &  0.96  &  0.33  &  05352+2247  &  CGCS 1006   &    \\
 05370+2629 N  &  181.43  &   $-$2.29  &  5.76  &  1.29  &  0.32  &  05371+2630  &  CGCS 1012   &    \\
 05416$-$4628 S  &  252.94  &  $-$30.71  &  1.13  &  0.95  &  0.27  &  05418$-$4628  &  CGCS 1052   &    \\
 05427+2040 N  &  187.07  &   $-$4.26  &  0.44  &  0.83  &  0.35  &  05426+2040  &  CGCS 1042   & Fig. 1 \\
 05465+1440 N  &  192.70  &   $-$6.59  &  4.63  &  1.16  &  0.33  &  05464+1440  &  CGCS 1076   &    \\
 05537+1158 N  &  195.93  &   $-$6.41  &  5.94  &  2.03  &  0.41  &  05537+1157  &              &    \\
 05555+1235 N  &  195.62  &   $-$5.74  &  4.57  &  1.07  &  0.30  &  05557+1236  &  CGCS 1124   &    \\
 06129$-$0526 N  &  213.72  &  $-$10.49  &  5.52  &  0.53  &  0.44  &  06129$-$0525  &              &    \\
 06179$-$0634 N  &  215.33  &   $-$9.89  &  6.15  &  0.83  &  0.38  &  06178$-$0634  &              &    \\
 06198$-$0558 N  &  215.00  &   $-$9.20  &  4.82  &  0.82  &  0.36  &  06196$-$0558  &  CGCS 1249   &    \\
 06256$-$1055 N  &  220.14  &  $-$10.12  &  5.75  &  0.69  &  0.26  &  06256$-$1055  &  CGCS 1283   &    \\
 06271$-$1400 N  &  223.12  &  $-$11.14  &  5.34  &  0.85  &  0.37  &  06272$-$1400  &  CGCS 1289   &    \\
 06465$-$2118 S  &  231.85  &  $-$10.13  &  4.55  &  0.69  &  0.36  &  06465$-$2118  &  CGCS 1416   &    \\
 06505$-$2257 S  &  233.77  &  $-$10.01  &  5.40  &  1.10  &  0.33  &  06504$-$2258  &              &    \\
 06553$-$2745 S  &  238.65  &  $-$11.10  &  5.98  &  0.78  &  0.44  &  06555$-$2744  &  CGCS 1490   &    \\
 06574$-$1440 S  &  227.01  &   $-$4.87  &  4.38  &  1.02  &  0.27  &  06575$-$1441  &  CGCS 1502   &    \\
 06589$-$1846 S  &  230.86  &   $-$6.40  &  6.07  &  0.46  &  0.44  &  06588$-$1847  &  CGCS 1513   &    \\
 07001$-$2412 S  &  235.88  &   $-$8.59  &  6.62  &  0.79  &  0.21  &  07000$-$2411  &  CGCS 1517   &    \\
 07018$-$1601 S  &  228.70  &   $-$4.54  &  4.47  &  0.78  &  0.39  &  07018$-$1601  &  CGCS 1532   &    \\
 07027$-$1456 S  &  227.84  &   $-$3.85  &  4.38  &  2.53  &  0.42  &  07028$-$1456  &  CGCS 1540   &    \\
 07030$-$1603 S  &  228.87  &   $-$4.30  &  5.09  &  1.43  &  0.37  &  07031$-$1604  &              & see note a \\
 07036$-$2221 S  &  234.57  &   $-$7.04  &  6.90  &  3.73  &  0.41  &  07036$-$2220  &              &    \\
 07037$-$2326 S  &  235.56  &   $-$7.51  &  6.56  &  0.93  &  0.26  &  07039$-$2326  &              & see note b \\
 07055$-$1441 S  &  227.94  &   $-$3.14  &  5.86  &  1.01  &  0.29  &  07054$-$1442  &  CGCS 1562   &    \\
 07060$-$1322 S  &  226.82  &   $-$2.42  &  5.24  &  0.71  &  0.34  &  07061$-$1323  &  CGCS 1567   &    \\
 07091$-$2005 S  &  233.12  &   $-$4.88  &  6.15  &  0.75  &  0.40  &  07091$-$2004  &  CGCS 1591   &    \\
 07099$-$2012 S  &  233.32  &   $-$4.76  &  4.03  &  2.89  &  0.42  &  07098$-$2012  &  CGCS 1601   &    \\
 07100$-$1102 S  &  225.22  &   $-$0.46  &  5.41  &  1.05  &  0.36  &  07100$-$1101  &  CGCS 1602   &    \\
 07104$-$1849 S  &  232.15  &   $-$4.00  &  6.18  &  0.93  &  0.29  &  07104$-$1849  &  CGCS 1606   &    \\
 07114$-$1525 S  &  229.25  &   $-$2.23  &  4.65  &  0.85  &  0.32  &  07113$-$1526  &  CGCS 1611   &    \\
 07116$-$1935 S  &  232.96  &   $-$4.13  &  3.32  &  0.88  &  0.32  &  07116$-$1936  &  CGCS 1612   &    \\
 07118$-$1023 S  &  224.83  &    0.22  &  5.42  &  2.25  &  0.45  &  07118$-$1022  &  CGCS 6209   &    \\
 07121$-$1155 S  &  226.23  &   $-$0.44  &  5.93  &  1.04  &  0.36  &  07120$-$1156  &              & see note c \\
 07124$-$1720 S  &  231.06  &   $-$2.90  &  3.66  &  0.74  &  0.35  &  07124$-$1720  &  CGCS 1615   &    \\
 07127$-$1030 S  &  225.05  &    0.37  &  4.88  &  0.70  &  0.39  &  07129$-$1030  &  CGCS 1618   &    \\
 07131$-$1728 S  &  231.25  &   $-$2.80  &  4.36  &  0.84  &  0.34  &  07132$-$1728  &  CGCS 1624   &    \\
 07136$-$1513 S  &  229.32  &   $-$1.65  &  5.65  &  1.15  &  0.27  &  07136$-$1512  &  CGCS 1630   &    \\
 07139$-$0907 S  &  223.96  &    1.28  &  5.87  &  1.10  &  0.24  &  07139$-$0907  &  CGCS 1631   &    \\
 07141$-$1442 S  &  228.92  &   $-$1.31  &  5.47  &  0.88  &  0.28  &  07141$-$1442  &  CGCS 1632   &    \\
 07145$-$1429 S  &  228.77  &   $-$1.11  &  4.53  &  1.90  &  0.32  &  07145$-$1428  &  CGCS 6214   &    \\
 07163$-$1608 S  &  230.43  &   $-$1.53  &  4.85  &  0.62  &  0.45  &  07163$-$1608  &  CGCS 1648   &    \\
 07189$-$0805 S  &  223.64  &    2.86  &  5.69  &  0.81  &  0.36  &  07189$-$0805  &  CGCS 1667   &    \\
 07193$-$1529 S  &  230.20  &   $-$0.57  &  4.85  &  1.06  &  0.25  &  07194$-$1529  &  CGCS 1672   &    \\
 07206$-$1032 S  &  225.99  &    2.05  &  4.16  &  0.99  &  0.30  &  07204$-$1032  &  CGCS 1682   &    \\
 07208$-$1411 S  &  229.23  &    0.37  &  3.76  &  0.95  &  0.30  &  07208$-$1410  &  CGCS 1687   &    \\
 07214$-$1525 S  &  230.39  &   $-$0.09  &  5.56  &  0.71  &  0.35  &  07215$-$1527  &  CGCS 1694   &    \\
 07216$-$1245 S  &  228.06  &    1.22  &  4.19  &  2.90  &  0.37  &  07217$-$1246  &  CGCS 1696   &    \\
 07222$-$1553 S  &  230.88  &   $-$0.16  &  5.31  &  1.12  &  0.15  &  07223$-$1553  &  CGCS 1701   &    \\
 07225$-$0538 S  &  221.89  &    4.80  &  6.26  &  2.23  &  0.29  &  07227$-$0537  &  CGCS 6223   &    \\
 07235$-$0354 S  &  220.49  &    5.84  &  3.64  &  0.98  &  0.28  &  07235$-$0356  &  CGCS 1708   & J-type \\
 07236$-$0505 S  &  221.54  &    5.31  &  4.75  &  0.75  &  0.42  &  07237$-$0505  &  CGCS 1709   &    \\
 07245$-$1108 S  &  226.98  &    2.61  &  4.39  &  0.86  &  0.25  &  07245$-$1107  &  CGCS 1719   &    \\
 07247$-$0901 S  &  225.15  &    3.67  &  4.48  &  1.85  &  0.31  &  07246$-$0903  &  CGCS 1720   &    \\
 07250$-$0832 S  &  224.75  &    3.96  &  5.18  &  0.84  &  0.32  &  07249$-$0832  &  CGCS 1724   &    \\
 07275$-$1152 S  &  227.97  &    2.91  &  5.38  &  0.88  &  0.31  &  07275$-$1152  &  CGCS 1733   &    \\
 07288$-$1007 S  &  226.60  &    4.03  &  4.41  &  0.89  &  0.35  &  07288$-$1006  &  CGCS 1743   &    \\
 07289$-$0823 S  &  225.08  &    4.89  &  5.12  &  0.78  &  0.34  &  07288$-$0824  &  CGCS 1742   &    \\
 07298$-$0415 S  &  221.53  &    7.06  &  5.55  &  0.74  &  0.35  &  07299$-$0415  &  CGCS 1755   &    \\
 07313$-$1149 S  &  228.38  &    3.74  &  6.31  &  0.61  &  0.45  &  07312$-$1148  &  CGCS 1767   & Fig. 2 \\
 07336$-$1005 S  &  227.14  &    5.07  &  7.05  &  3.84  &  0.40  &  07336$-$1006  &  CGCS 6233   & see note d \\
 07486$-$0229 S  &  222.24  &   12.03  &  3.00  &  1.76  &  0.32  &  07487$-$0229  &  CGCS 1907   &  \\
 11318$-$7256 N  &  297.33  &  $-$11.20  &  1.32  &  1.94  &  0.36  &  11318$-$7256  &  CGCS 3062   & see note d \\
 11372$-$7217 N  &  297.54  &  $-$10.45  &  2.01  &  0.51  &  0.39  &  11371$-$7216  &  CGCS 3083   &    \\
 12545+6615 S  &  122.12  &   51.14  &  0.24  &  0.88  &  0.36  &  12544+6615  &  CGCS 3313   & J-type   \\
 13411$-$7021 N  &  307.39  &   $-$8.21  &  2.62  &  0.98  &  0.28  &  13408$-$7021  &  CGCS 3425   &    \\
 16010$-$5837 N  &  325.85  &   $-$4.80  &  3.57  &  0.44  &  0.36  &  16009$-$5837  &  CGCS 6615   &    \\
 16522$-$4616 N  &  340.13  &   $-$1.91  &  3.61  &  1.07  &  0.34  &  16522$-$4616  &  CGCS 3745   &    \\
 17055$-$4343 N  &  343.61  &   $-$2.19  &  3.41  &  1.25  &  0.30  &  17054$-$4342  &  CGCS 3773   &    \\
 17131$-$3906 N  &  348.16  &   $-$0.60  &  4.45  &  2.51  &  0.39  &  17130$-$3907  &  CGCS 3794   &    \\
 17145$-$3431 N  &  352.05  &    1.84  &  3.70  &  0.89  &  0.42  &  17146$-$3432  &  CGCS 3798   &    \\
 17173$-$4019 N  &  347.64  &   $-$1.97  &  1.38  &  0.99  &  0.28  &  17172$-$4020  &  CGCS 3808   &    \\
 17181$-$3812 N  &  349.47  &   $-$0.88  &  3.90  &  1.45  &  0.43  &  17183$-$3814  &              & see note e \\
 17233$-$3132 N  &  355.56  &    2.03  &  3.90  &  1.27  &  0.28  &  17233$-$3132  &  CGCS 3830   &    \\
 17420$-$1836 N  &    8.74  &    5.48  &  2.23  &  0.62  &  0.44  &  17419$-$1838  &  CGCS 3875   &    \\
 18306+3657 S  &   65.35  &   19.47  &  0.26  &  0.96  &  0.17  &  18306+3657  &  CGCS 4038   & J-type  \\ 
 19028+2018 S  &   52.66  &    6.28  &  4.65  &  2.93  &  0.36  &  19029+2017  &  CGCS 6767   &    \\
 19085+1329 S  &   47.21  &    1.94  &  4.54  &  0.98  &  0.44  &  19085+1327  &  CGCS 4192   &    \\
 19109+1156 S  &   46.10  &    0.71  &  3.84  &  2.06  &  0.41  &  19108+1155  &  CGCS 4202   &    \\
 19119+1022 S  &   44.83  &   $-$0.25  &  4.52  &  1.45  &  0.44  &  19118+1020  &            & see note f \\
 19121+1830 S  &   52.04  &    3.51  &  3.99  &  1.06  &  0.32  &  19122+1830  &  CGCS 4209   &    \\
 19131+1718 S  &   51.11  &    2.73  &  3.60  &  0.97  &  0.30  &  19132+1719  &  CGCS 4211   &    \\
 19155+0846 S  &   43.84  &   $-$1.80  &  3.71  &  1.09  &  0.30  &  19155+0847  &  CGCS 4224   &    \\
 19181+1250 S  &   47.73  &   $-$0.44  &  4.03  &  0.99  &  0.34  &  19181+1249  &  CGCS 4238   &    \\
 19188+0735 S  &   43.18  &   $-$3.06  &  4.95  &  0.93  &  0.36  &  19188+0733  &  CGCS 4240   &    \\
 19207+1719 S  &   51.97  &    1.15  &  4.34  &  1.73  &  0.42  &  19207+1720  &  CGCS 4251   &    \\
 19218+4607 N  &   77.99  &   14.12  &  5.07  &  0.65  &  0.40  &  19217+4607  &  CGCS 4260   &    \\
 19225+1352 S  &   49.16  &   $-$0.88  &  4.99  &  0.76  &  0.36  &  19227+1354  &  CGCS 4261   &    \\
 19248+0656 S  &   43.31  &   $-$4.67  &  3.48  &  2.43  &  0.43  &  19248+0658  &  CGCS 4275   & Fig. 3 \\
 19284+1340 S  &   49.66  &   $-$2.23  &  4.55  &  1.02  &  0.28  &  19285+1340  &  CGCS 4312   &    \\
 19353+0636 S  &   44.27  &   $-$7.15  &  3.25  &  0.87  &  0.26  &  19354+0636  &  CGCS 4373   &    \\
 19505+5332 N  &   86.93  &   13.31  &  5.00  &  0.77  &  0.36  &  19505+5333  &  CGCS 4543   & Fig. 5 \\
 20229$-$1641 S  &   27.84  &  $-$28.12  &  6.57  &  1.04  &  0.27  &              &  CGCS 4844   &    \\
 20547+6404 N  &  100.56  &   12.21  &  4.70  &  2.31  &  0.41  &  20546+6405  &  CGCS 5053   &    \\
 21263+6959 N  &  107.19  &   13.89  &  1.47  &  1.37  &  0.37  &  21262+7000  &  CGCS 5322   &    \\
 21538+6916 N  &  108.51  &   11.74  &  5.03  &  0.74  &  0.39  &  21539+6916  &  CGCS 5507   &    \\
 21563+7109 N  &  109.89  &   13.08  &  3.64  &  0.72  &  0.39  &  21560+7108  &  CGCS 5529   &    \\
\tableline
05590+0637 N &  201.27 & $-$7.91  &  4.34 &  1.08  & 0.46  &  05591+0638 & CGCS 1145 & see note g\\
07175$-$0843 S &  224.03 &  2.25  &  5.69 &  0.81  & 0.47  &  07175$-$0842 & CGCS 1655 & see note g\\
07289$-$0252 S &  220.20 &  7.50  &  5.93 &  0.73  & 0.47  &  07290$-$0251 & CGCS 1744 & see note g\\
\tableline
07281$-$1216 S &  228.39 &  2.84  &  5.83 &  0.61  & 0.46  &  07280$-$1217 & CGCS 1735 & see note h \\
\enddata



\tablenotetext{a}{A 30-min-exposure plate obtained by D.J. MacConnell shows 
nothing at this position. The R DSS2 plate has a faint star 3\arcsec ~SE of 
the IRAS position}
\tablenotetext{b}{A 30-min-exposure plate obtained by D.J. MacConnell shows 
a faint continuum (short and red) which has a general character of a carbon 
star but is too underexposed to be certain, although it has been tentatively 
identified as a C star by MacConnell (1979). 
The R DSS2 plate has a bright star at the IRAS position}
\tablenotetext{c}{A 60-min-exposure plate obtained by D.J. MacConnell shows 
nothing at the IRAS position}
\tablenotetext{d}{2.9--3.5 $\mu$m spectrum also presented by Groenewegen 
et al. (1994)}
\tablenotetext{e}{A 60-min-exposure plate obtained by D.J. MacConnell has a 
very red, non-C star east of the IRAS position which probably corresponds to 
the bright star 32\arcsec ~away at position angle 66 deg. on the DDS2 I plate}
\tablenotetext{f}{{LRS} spectrum C (11 $\mu$m SiC emission, Kwok et al. 1997); 
tentative identification as a C star by Chen \& Chen (2003)}
\tablenotetext{g}{3.1 $\mu$m index larger than 0.45 (paper II)}
\tablenotetext{h}{3.1 $\mu$m index larger than 0.45 (Tanaka et al., in preparation)}

\end{deluxetable}


\end{document}